\documentclass[preprint,showpacs,preprintnumbers,amsmath,amssymb]{revtex4}%
\usepackage{graphicx}% Include figure files
\usepackage{dcolumn}
\usepackage{bm}
\usepackage{amsmath}
\usepackage{amsfonts}
\usepackage{amssymb}%
\providecommand{\U}[1]{\protect\rule{.1in}{.1in}}
\newcommand{\be}{\begin{equation}}
\newcommand{\en}{\end{equation}}
\newcommand{\bea}{\begin{eqnarray}}
\newcommand{\ena}{\end{eqnarray}}
\begin{document}
\title{General
dissipative coefficient in warm Intermediate and logamediate
inflation}
\author{Ram\'on Herrera}
\email{ramon.herrera@ucv.cl}
\author{Marco Olivares}
\email{marco.olivares@ucv.cl}
\author{Nelson Videla}
\email{nelson.videla@ucv.cl}
\affiliation{Instituto de F\'{\i}sica, Pontificia Universidad Cat\'{o}lica de
Valpara\'{\i}so, Avenida Brasil 2950, Casilla 4059, Valpara\'{\i}so, Chile.}
\date{\today}

\begin{abstract}
We study a general form for the dissipative coefficient
$\Gamma(T,\phi)=C_\phi\,T^{m}/\phi^{m-1}$ in the context of warm
intermediate and logamediate inflationary universe models. We
analyze these models in the weak and strong dissipative regimes.
In the slow-roll approximation, we describe in great detail the
characteristics of these models. In both regimes, we use recent
data from the WMAP nine-year data and Planck data to constrain
the parameters appearing in our models.
\end{abstract}

\maketitle

%\preprint{GACG/07/2006}

%PACS, the Physics and Astronomy
%Classification Scheme.
%\keywords{Suggested keywords}%Use showkeys class option if keyword
%display desired

\section{Introduction}

It is well known that the inflationary universe model provides an
interesting approach for solving some of the problems of the standard
big bang model, such as the flatness, the horizon
etc.\cite{R1,R102,R103,R104,R105,R106}. One of the achievements of the
inflation  scenario is that it scenario can  offer an elegant mechanism to
explain the large-scale structure \cite{R2,R202,R203,R204,R205}
and the observed anisotropy of the cosmic microwave background
(CMB) radiation\cite{astro,astro2,astro202,Planck}.

On the other hand, the warm inflation scenario differs from
cold inflation  in that there is no separate reheating period  in
the former; rather, radiation production occurs at the same
time as inflationary expansion due to the decay of the
inflaton field into radiation and particles during the slow-roll
phase.\cite{warm}. In this form,  the universe ceases inflating
and smoothly enters a radiation-dominated big bang scenario
\cite{warm,taylorberera,taylorberera02,taylorberera03,taylorberera04,taylorberera05,
taylorberera06,taylorberera07,taylorberera08,taylorberera09}. In
the inflationary regime, the dissipative effects are important and originate
 from a friction term which describes the physical
processes of the scalar field dissipating into a thermal bath.
Also, during warm inflation the thermal fluctuations represent a
dominant role in producing the initial density fluctuations
necessary for large-scale Structure (LSS) formation. Here, these
density fluctuations arise from thermal rather than quantum
fluctuations \cite{62526,6252602,6252603,6252604,1126}.

In the context of the dissipation coefficient $\Gamma$,
the particular scenario of low-temperature regimes was presented in
Refs.\cite{26,28,2802}. There, the value of $\Gamma$ was
considered in supersymmetric models which have an inflaton
together with multiplets of heavy and light fields. Different
choices  of $m$ have been adopted or, equivalently, different
expressions for the dissipation coefficient have been analyzed in
Refs.\cite{Zhang:2009ge,BasteroGil:2011xd,BasteroGil:2012cm}. In
this form, following Refs.\cite{Zhang:2009ge,BasteroGil:2011xd},
we consider a general form of the dissipative coefficient, given
by

\begin{equation}
\Gamma=C_{\phi}\,\frac{T^{m}}{\phi^{m-1}}, \label{G}%
\end{equation}
where $m$ is an integer and $C_\phi$ is associated to the
dissipative microscopic dynamics. In particular, for the case
$m=3$, the value of $C_\phi$ corresponds to
   $C_{\phi}=0.64\,h^{4}\,\mathcal{N}$, in which ${\mathcal{N}}%
={\mathcal{N}}_{\chi}{\mathcal{N}}_{decay}^{2}$, where
$\mathcal{N}_{\chi}$ is the multiplicity of the $X$ superfield and
${\mathcal{N}}_{decay}$ is the number of decay channels available
in $X$'s decay\cite{26,27} (see also
Refs.\cite{Berera:2008ar,BasteroGil:2010pb}). The dissipation
coefficient $\Gamma$ for $m=1$ is given by $\Gamma\propto\,T$ and
it represents  the high-temperature supersymmetry (SUSY) case;
when $m=0$ the dissipation coefficient is $\Gamma\propto\phi$ and
it corresponds to an exponentially decaying propagator in the SUSY
case; and  when $m=-1$ then $\Gamma\propto\phi^2/T$, and it
corresponds to the non-SUSY case. In particular, the case $m=3$,
i.e., $\Gamma\propto T^{3}/{\phi^2}$ was considered for the warm intermediate model
 in Ref.\cite{delCampo:2009xi} and for the
warm logamediate model in Ref.\cite{Herrera:2012zz}.
In the following, we will study the warm intermediate and
logamediate inflationary models for the values of $m=1$, $m=0$, and
$m=-1$.

On the other hand, the most interesting exact solutions in the inflationary universe
 is can be found  by using an exponential potential, which is
often called a power-law inflation since the scale factor has a
 power-law-type evolution, i.e., $a(t)\sim t^{p}$, in which
$p>1$\cite{power}. Also, an exact solution can be obtained in the
de Sitter inflationary universe, where a constant scalar potential
is considered; see Ref.\cite{R1}. However,  exact solutions can
also be obtained for the scenario of intermediate inflation
\cite{Barrow1}. In this universe model the scale factor
$a(t)$ growths as
\begin{equation}
a(t)=\exp[\,A\,t^{f}],  \label{at}
\end{equation}
where $A$ and $f$ are two constants; $A>0$ and $0<f<1$
\cite{Barrow1}. This expansion type  is slower than de Sitter
inflation, but faster than power-law inflation; this is why it
is known as "intermediate". Nevertheless, a generalized model of
the expansion of the universe is logamediate inflation \cite{R12}.
In this model, the scale factor $a(t)$ increases as
\begin{equation} a(t)=\exp[\,A\,(\ln t)^{\lambda}],\label{at1}
\end{equation}
where  $\lambda$ and $A$ are dimensionless constant parameters
such that $\lambda > 1$ and $A > 0$; see Ref.\cite{R12}. Note
that  for special the case in which $\lambda=1$ and $A=p$, the
logamediate inflation model becomes a power-law
inflation model \cite{power}.

Intermediate and logamediate models were originally developed  as
an exact solution, but they may be best formulated  from the
slow-roll approximation. During the slow-roll approximation, it is
possible to find a spectral index $n_s\sim 1$. In particular, in
the model of intermediate inflation  the value $n_s=1$ that
corresponds to the Harrizon-Zel'dovich spectrum is found  for the
special value $f=2/3$ \cite{Barrow2}, but this value is
not suported  by the current  observational
data\cite{astro,astro2,Planck}. Also, an important observational
quantity in both  models, is that the tensor-to-scalar ratio $r$, which
is significantly $r\neq 0$\cite{ratior,Barrow3}.

%The motivation to study logamediate inflationary universe models
%becomes from the form of the effective potential  that arise with
%this scale factor.

%%%%%%%%%%%%%%%%%%%%%%%%%%%%%%%%%%%%%%%%%%%%%%%

The main goal of the present work is to analyze the possible
realization of an expanding  intermediate and logamediate scale
factor within the framework of a warm inflationary universe model,
and how warm intermediate and logamediate inflation works  with  a
generalized form of the dissipative coefficient. We will study these
models for two regimes: the weak and the strong dissipative
scenarios.
%{\bf In this context  we would like to introduce the warm
%inflation as a mechanism to bring logamediate inflation to an end.
%Therefore, the main goal of the present paper is to implement
%dissipative effects  into the logamediate inflationary scenario
%and see what consequences we may extract.}

The  paper is organized as follows. The next section presents the
basic equations for warm inflation. In Secs. III and IV, we
discuss the weak and strong dissipative regimes in the
intermediate and logamediate scenarios. In both sections, we give
explicit expressions for the dissipative coefficient, the scalar
potential, the scalar power spectrum and the tensor-to-scalar ratio
for these models. The nine-year WMAP data  and Planck data are
used to constrain the parameters in our models.  In Sec. V, we
analyze the interpolation between the weak and strong decays.
Finally, Sec. VI summarizes and discuses our finding. We use
units in which $c=\hbar=1$.

\section{Warm Inflation: Basic equations\label{secti}}

We consider a spatially  flat Friedmann equation, given by
\begin{equation}
H^{2}=\frac{\kappa}{3}\,\rho=\frac{\kappa}{3}\,[\rho_{\phi}+\rho_{\gamma}],
\label{HC}%
\end{equation}
where $H$ denotes the Hubble parameter given by $H=\dot{a}/a$,
$a$ is the scale factor, and the constant $\kappa=8\pi
G=8\pi/m_{p}^{2}$, where $m_{p}$ denotes the Planck mass. In warm
inflation, the universe is filled with a self-interacting scalar
field with energy density $\rho_{\phi}$ and a radiation field with
energy density $\rho_{\gamma}$.  Here, the total energy density
$\rho$ is given by $\rho=\rho_{\phi}+\rho_{\gamma}$.

In the following, we will consider that the energy density related
to the scalar field is given by
$\rho_{\phi}=\dot{\phi}^{2}/2-V(\phi)$ and the pressure is
$P_{\phi}=\dot{\phi}^{2}/2+V(\phi)$, where $V(\phi)$ corresponds
to the effective  potential. Dots mean derivatives with respect to
time.

The dynamical equations for $\rho_{\phi}$ and $\rho_{\gamma}$ in
warm inflation are described by\cite{warm}
\begin{equation}
\dot{\rho_{\phi}}+3\,H\,(\rho_{\phi}+P_{\phi})=-\Gamma\;\;\dot{\phi}^{2},
\label{key_01}%
\end{equation}
and
\begin{equation}
\dot{\rho}_{\gamma}+4H\rho_{\gamma}=\Gamma\dot{\phi}^{2}, \label{key_02}%
\end{equation}
where $\Gamma>0$ is the dissipation coefficient and it is
responsible for the decay of the scalar field into radiation during
the inflationary scenario. The dissipation coefficient $\Gamma$
can be considered to be a function of the temperature of the
thermal bath $\Gamma(T)$, the scalar field $\Gamma(\phi)$,
both $\Gamma(T,\phi)$, or a constant\cite{warm}.

During warm inflation $\rho_\phi\gg\rho_\gamma$, i.e., the energy
density related to the scalar field predominates  over the energy
density of the radiation field. Then Eq.(\ref{HC})
becomes\cite{warm,62526,6252602,6252603,6252604}
\begin{equation}
H^{2}\approx\frac{\kappa}{3}\,\rho_{\phi}=\frac{\kappa}{3}\,\left[\frac{\dot{\phi}}{2}+V(\phi)\right]. \label{inf2}%
\end{equation}

Combining Eqs. (\ref{key_01}) and (\ref{inf2}) yields
\begin{equation}
\dot{\phi}^{2}= {\frac{2 }{\kappa}}\frac{-\dot{H}}{(1+R)}, \label{inf3}%
\end{equation}
where $R$ denotes the rate between $\Gamma$ and the Hubble
parameter, i.e., $R=\frac{\Gamma}{3H}$. Here, we note that for the
the weak or strong dissipation regime, the rate is $R<1$ or
$R>1$, respectively.

Following Refs.\cite{warm,62526,6252602,6252603,6252604}, we
consider   that during warm inflation the radiation production is
quasistable, in which  $\dot{\rho
}_{\gamma}\ll4H\rho_{\gamma}$ and $\dot{\rho}_{\gamma}\ll\Gamma\dot{\phi}^{2}%
$. In this form, by combing Eqs.(\ref{key_02}) and (\ref{inf3}) we
get
\begin{equation}
\rho_{\gamma}=\frac{\Gamma\dot{\phi}^{2}}{4H}=\frac{\Gamma\,(-\dot{H}%
)}{2\,\kappa\,H\,(1+R)}. \label{rh}%
\end{equation}

On the other hand, the energy density of the radiation field
$\rho_\gamma$
could be written as $\rho_{\gamma}=C_{\gamma}\,T^{4}$, where  $C_{\gamma}%
=\pi^{2}\,g_{\ast}/30$. Here,  $g_{\ast}$ corresponds to the
number of relativistic degrees of freedom. Combining the above
relation for the energy density $\rho_\gamma$  and
Eq.(\ref{rh}), we get that the temperature of the thermal bath is
\begin{equation} T=\left[
\frac{\Gamma\,(-\dot{H})}{2\,\kappa\,\,C_{\gamma}H\,(1+R)}\right]
^{1/4}. \label{rh-1}%
\end{equation}

In this form,  Eqs.(\ref{G}) and (\ref{rh-1}) combine to become
\begin{equation}
\Gamma^{{\frac{4-m }{4}}}=\,\alpha_{m}\phi^{1-m} \left[
\frac{-\dot{H}}{H}\right]  ^{m/4}\,(1+R)^{-m/4}, \label{G1}%
\end{equation}
where the constant $\alpha_{m}$ is given by $\alpha_{m}=
C_{\phi}\left[ \frac{1}{2\kappa\, C_{\gamma}}\right] ^{m/4} $.
 Here, we note that the expression given by Eq.(\ref{G1})
specifies the dissipation coefficient in the weak dissipative
regime, in which $\Gamma^{{\frac{4-m }{4}}}=\,\alpha_{m}\phi^{1-m}
\left[ \frac{-\dot{H}}{H}\right]  ^{m/4}\,$, or in the strong
dissipative regime, where $\Gamma=\,\alpha_{m}\phi^{1-m} \left[
-3\,\dot{H}\right] ^{m/4}\,$.

On the other hand, from Eqs.(\ref{HC}), (\ref{inf3}), and
(\ref{rh}),
the effective potential becomes%

\begin{equation}
V=\frac{3}{\kappa}\,H^{2}+\frac{\dot{H}}{\kappa(1+R)}\,\left(
1+\frac{3}{2}\,R\right)  , \label{pot}%
\end{equation}
which also could be calculated  explicitly in terms of the scalar
field $\phi$, i.e., $V=V(\phi)$.

In the following, we will analyze the intermediate and logamediate
models in the context of warm inflation for a general form of the
dissipative coefficient $\Gamma(T,\phi)=C_\phi\,T^{m}/\phi^{m-1}$
for the values $m=1$, $m=0$, and $m=-1$. Also, we will restrict
ourselves to the weak (or strong ) dissipative regime. We recall
that the case $m=3$ was considered for the warm intermedite model
in Ref.\cite{delCampo:2009xi} and for the warm logamediate model
 in Ref.\cite{Herrera:2012zz}.

\section{ The weak dissipative regime $\Gamma<3H$\label{section3}}

\subsection{ Warm intermediate inflation\label{subsection1}}

Considering that our warm model evolves according to the weak
dissipative regime, in which $\Gamma<3H$, and combining
Eqs.(\ref{at}) and (\ref{inf3}), we get
\begin{equation}
\phi(t)-\phi_0=k_{0}\,t^{f/2}, \label{wr1}%
\end{equation}
where $k_{0}\equiv\sqrt{\frac{8\,A\,(1-f)}{\kappa f}}$ is a
constant and $\phi(t=0)=\phi_0$ is an integration constant that
without loss of generality can be taken as $\phi_{0}=0$. From
Eq.(\ref{wr1}), the Hubble parameter as a function of the inflaton
field gives $ H(\phi) =A\,f\,\left( \frac{k_{0}}{\phi}\right)
^{{\frac{2(1-f)}{f}} }\propto\,\phi^{2(f-1)/f}.$

Considering Eq.(\ref{pot}), the effective potential in the weak
dissipative regime becomes
\begin{equation}
V(\phi)=k_{1}\phi^{-\beta_{1}}, \label{pot11}%
\end{equation}
where the constants $k_1$ and $\beta_1$ are given by
$k_{1}=\frac{3}{\kappa}(A\,f)^{2}k_{0}^{\beta_{1}},$ and $\beta
_{1}={\frac{4(1-f)}{f}}$. Note that this kind of
scalar potential given by Eq.(\ref{pot11}) coincides with the
effective potential calculated  in Ref.\cite{R12}. Since
$R=\Gamma/3H<1$, then from Eqs.(\ref{G1}) and (\ref{wr1}) the
dissipation coefficient $\Gamma$ in terms of the scalar field
becomes
\begin{equation}
\Gamma(\phi)=k_{2}\,\phi^{\,\beta_{2}}, \label{gammaph}%
\end{equation}
where $k_{2}=C_{\phi}^{{\frac{4}{4-m}}}\left[  \frac{(1-f)k_{0}^{2/f}}%
{2\kappa\,C_{\gamma}}\right]  ^{\frac{m}{4-m}}$ and
$\beta_{2}={\frac {4f(1-m)-2m}{f(4-m)}}$.

On the other hand, the dimensionless slow-roll parameter
$\varepsilon$ is given by $
\varepsilon=-\frac{\dot{H}}{H^{2}}=k_{0}^{2}\left(
{\frac{1-f}{Af}}\right) \phi^{-2} $, and then the condition for
inflation to occur, $\ddot{a}>0$ (or equivalently
$\varepsilon<$1), is  satisfied when $\phi>k_{0}\left(  {\frac{1-f}{Af}%
}\right)  ^{1/2}$. Following Ref.\cite{R12}, the  inflationary
phase  begins at the earliest possible stage, that is, at
$\varepsilon=1$, and hence the scalar field $\phi_{1}$ can be
expressed as $ \phi_{1}=k_{0}\left( {\frac{1-f}{Af}}\right)
^{1/2}. $ Also, from Eq.(\ref{wr1}) the number of $e$-folds $N$
between two different values of cosmological times $t_{1}$ and
$t_{2}$ or equalivalently  between  $\phi_{1}$ and $\phi_{2}$ is given
by
\begin{equation}
N=\int_{t_{1}}^{t_{2}}\,H\,dt=A\,\left(  t_{2}^{f}-t_{1}^{f}\right)
=A\,k_{0}^{-2}\,\left(  \phi_{2}^{2}-\phi_{1}^{2}\right)  . \label{N1}%
\end{equation}

In the following, we will describe the scalar and tensor
perturbations  for our warm model in the weak dissipative regime.
For a standard scalar field the density perturbation could be written as ${\mathcal{P}%
_{\mathcal{R}}}^{1/2}=\frac{H}{\dot{\phi}}\,\delta\phi$\cite{warm}.
During the warm inflation, a thermalized radiation component is
present and the fluctuations $\delta\phi$ are dominantly thermal
rather than quantum\cite{warm,62526,6252602,6252603,6252604}.
Following Refs.\cite{62526,6252602,6252603,6252604,B1}, in the weak
dissipative regime, the value of $\delta\phi^{2}$ is given by
$\delta\phi^{2}\simeq H\,T$. In this form, by combining
Eqs.(\ref{inf3}), (\ref{rh-1})
and (\ref{G1}) the power spectrum of the scalar perturbation ${\mathcal{P}_{\mathcal{R}}}$ yields%
\begin{equation}
{\mathcal{P}_{\mathcal{R}}}=\frac{\kappa}{2}\left(  \frac{C_{\phi}}{2\kappa
C_{\gamma}}\right)  ^{{\frac{1}{4-m}}}\phi^{{\frac{1-m}{4-m}}}H^{{\frac
{11-3m}{4-m}}}(-\dot{H})^{-{\frac{(3-m)}{4-m}}}. \label{pd}%
\end{equation}

From  Eqs.(\ref{wr1}) and (\ref{pd}) we get the power
spectrum as a function of the field,
\begin{equation}
{\mathcal{P}_{\mathcal{R}}}=k_{3}\,\phi^{-\beta_{3}}, \label{pd3}%
\end{equation}
where the constants $k_{3}$ and $\beta_{3}$ are defined as $
k_{3}=\frac{\kappa}{2}\left(  \frac{C_{\phi}}{2\kappa
C_{\gamma}}\right)
^{{\frac{1}{4-m}}}k_{0}^{{\frac{1-m}{4-m}}+\beta_{3}}A^{2}\,f^{2}%
\,(1-f)^{-{\frac{(3-m)}{4-m}}}$ and $
\beta_{3}={\frac{10-2m-f(17-5m)}{f(4-m)}}$.

The scalar spectral index $n_{s}$ is given by  $n_{s}-1=\frac{d\ln
\,{\mathcal{P}_{R}}}{d\ln k}$. Using Eqs. (\ref{wr1}) and
(\ref{pd}), the scalar spectral index $n_s$ is
\begin{equation}
n_{s}=1-\frac{4(1-f)[10-2m-f(17-5m)]}{\kappa
f^{2}(4-m)}\,\,\phi^{-2}.
\label{nss1}%
\end{equation}
Also, the spectral index $n_{s}$ can be written in terms of the
number of $e$-folds $N$. In this way, combining Eqs.(\ref{N1}) and
(\ref{nss1}), gives
\begin{equation}
n_{s}=1-\frac{10-2m-f(17-5m)}{2(4-m)[1+f(N-1)]}. \label{nswr}%
\end{equation}

Note that we can express the value of $f$ in terms of  $m$,
$n_{s}$, and $N$ as $
f=\frac{10-2m-2(4-m)(1-n_{s})}{17-5m+2(4-m)(1-n_{s})(N-1)}. $ In
particular, for the values $m=1$, $n_s=0.96$, and $N=60$ we get
that the value $f\simeq 0.30$ for $m=0$ corresponds to $f\simeq
0.27$, and for $m=-1$ it corresponds to $f\simeq 0.25$.

From Eqs.(\ref{N1}) and (\ref{pd3}) we can also express the value
of the parameter $A$ in terms of the parameters $C_{\phi}$,$C_{\gamma}$,${\mathcal{P}%
_{\mathcal{R}}}$, $m$, $n_{s}$, and $N$ as

\begin{equation}
A=k_{4}\left(  {\frac{2}{\kappa}}\right)  ^{{\frac{f(4-m)}{5-m}}}\left(
\frac{2\kappa C_{\gamma}}{C_{\phi}}\right)  ^{{\frac{f}{5-m}}}%
[1+f(N-1)]^{{\frac{10-2m-f(17-5m)}{2(5-m)}}}{\mathcal{P}_{\mathcal{R}}%
}^{{\frac{f(4-m)}{5-m}}}, \label{awr}%
\end{equation}
where the constant $k_{4}$ is given by
$
k_{4}=\left[  \frac{\kappa}{8(1-f)}\right]  ^{{\frac{f(1-m)}{2(5-m)}}%
}f^{-(1+f{\frac{m-1}{5-m}})}(1-f)^{{\frac{f(3-m)}{5-m}}}.$

Also, we can obtain an expression for the rate $R=\Gamma/3H$ in
terms of the scalar spectral index $n_{s}$. Considering
Eqs.(\ref{gammaph}) and (\ref{nss1}), we get
\begin{equation}
R(n_{s})=\frac{k_{2}}{3Afk_{0}^{\frac{2(1-f)}{f}}}\left[  \frac
{4(1-f)(10-2m-f(17-5m)}{f^{2}(4-m)(1-n_{s})}\right]  ^{\frac{2(2-m)-f(m+2)}%
{f(4-m)}}. \label{Ratio-ns}%
\end{equation}
\qquad

On the other hand, the generation of tensor perturbations during the
inflationary scenario would generate gravitational waves; see
Ref.\cite{Bha}. The spectrum of the tensor perturbations
${\mathcal{P}}_{g}$ is given by
${\mathcal{P}}_{g}=8\kappa(H/2\pi)^{2}$. An important observational
quantity is the tensor-to-scalar ratio
$r=\left(\frac{{\mathcal{P}}_g}{P_{\mathcal{R}}}\right)$. From
Eq.(\ref{pd}) we may write the tensor-to-scalar ratio $r$ in the regime
$R<1$ as $ r(k)=\left(
\frac{{\mathcal{P}}_{g}}{P_{\mathcal{R}}}\right)  \simeq
k_{5}\;\phi^{\beta_{5}}, $
where the constants $k_{5}={\frac{1 }{k_{3}}}\left(  {\frac{2 \kappa A^{2} f^{2} }{\pi^{2}}%
}\right)  k_{0}^{{\frac{4(1-f) }{f}}} $ and
$\beta_{5}=\beta_{3}-4(1-f)/f$. Also, the tensor-to-scalar ratio
can be rewritten in terms of the scalar spectral index as
\begin{equation}
r \simeq k_{5}\,k_{0}^{\beta_{5}}\left[  {\frac{10-2m-f(17-5m)}%
{2Af(4-m)(1-n_{s})}}\right]  ^{{\frac{\beta_{5} }{2}}}. \label{Rk11}%
\end{equation}

Analogously, as before the tensor-to-scalar ratio as a function of the
number of $e$-foldings $N$ can be written as $ r \simeq
k_{5}\,k_{0}^{\beta_{5}}\left[ {\frac{1+f(N-1)}{Af}}\right]
^{{\frac{\beta_{5} }{2}}}. $

\begin{figure}[th]
\includegraphics[width=2.5in,angle=0,clip=true]{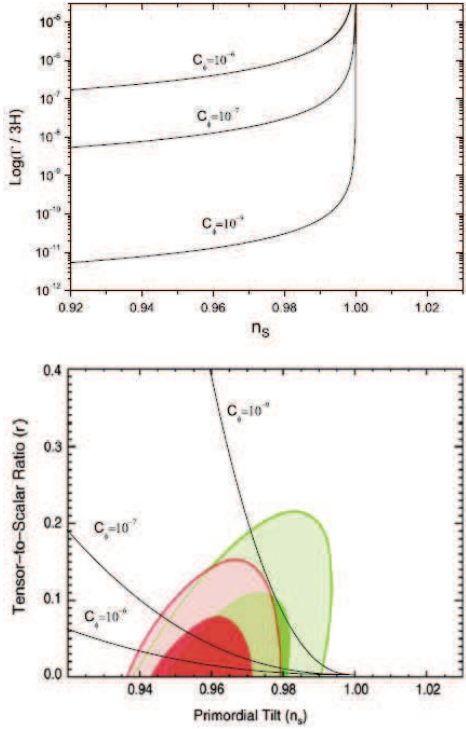}
{\hspace{4.cm}}
\caption{The evolution of the ratio
$R=\Gamma/3H$ versus the primordial tilt $n_s$ (upper panel) and
the evolution of the tensor-to-scalar ratio r versus $
n_s$ (lower panel) in the  warm intermediate weak dissipative regime for the
case $m=1$ i.e., $\Gamma\propto T$. In both panels we use three
different values of the parameter $C_\phi$, and
$\kappa=1$ and $C_\gamma=70$. In the lower panel, we show the
two-dimensional marginalized constraints (68$\%$ and 95$\%$
C.L.) on the inflationary parameters $r$ and $n_s$,
derived with the nine-year WMAP in conjunction with eCMB (green)
and eCMB+BAO+$H_0$ (red); see Ref.\cite{astro202}. \label{fig1}}
\end{figure}

%\begin{figure}[th]
%\includegraphics[width=3.7in,angle=0,clip=true]{fig4.eps}

%\hspace*{-4.0cm}{\vspace{-3.0cm}}{\includegraphics[width=3.6in,angle=0,clip=true]{planck1.eps}}
%\caption{Evolution of the tensor-scalar ratio $r$  versus the
%scalar spectrum index $n_s$  in the weak dissipative regime, for
%different values of the parameter $C_\phi$. In the upper panel, we
%have take $m=0$ ($\Gamma\propto\phi$)and  the lower panel, $m=-1$
%($\Gamma\propto\phi^2/T$). In both panels we used $\kappa=1$ and
%$C_\gamma=70$. The nine-year WMAP data places stronger limits on
%$r$ versus $n_s$\cite{astro202}. \label{fig2}}
%\end{figure}

\begin{figure}[th]
\includegraphics[width=2.5in,angle=0,clip=true]{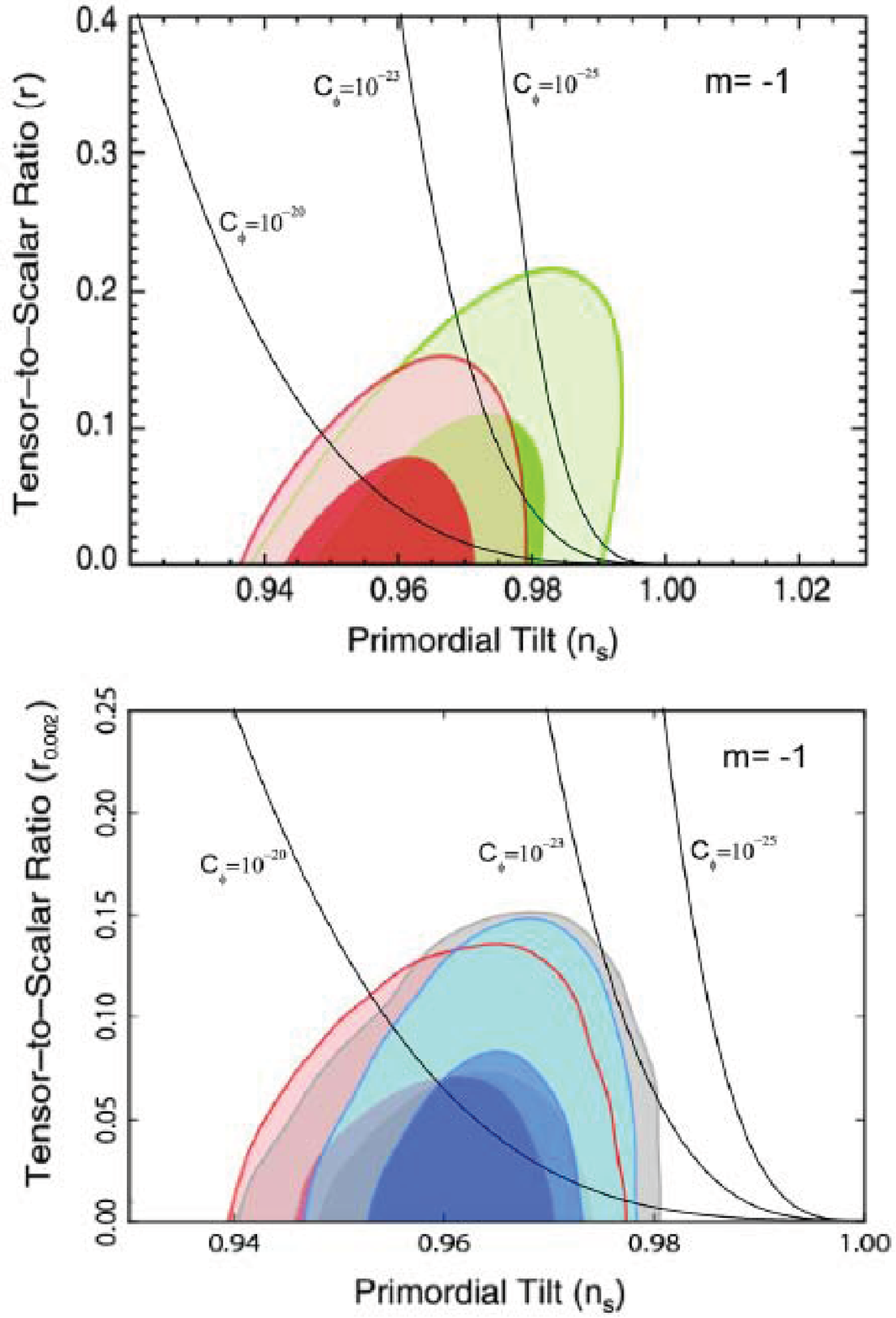}

%{\hspace{-2.5cm}\includegraphics[width=3.55in,angle=0,clip=true]{planck1.eps}}
\caption{Evolution of the tensor-to-scalar ratio $r$  versus the
scalar spectrum index $n_s$  in the warm intermediate weak
dissipative regime, for different values of the parameter $C_\phi$
and $m=-1$ i.e., $\Gamma\propto\phi^2/T$. In the upper panel we show the
two-dimensional marginalized constraints (68$\%$ and 95$\%$
C.L.) on the inflationary parameters $r$ and $n_s$
derived with the nine-year WMAP \cite{astro202}. In the lower
panel we show the constraints from the Planck combination with other data sets \cite{Planck}.
In both panels $\kappa=1$ and $C_\gamma=70$. \label{fig2}}
\end{figure}

In Fig.\ref{fig1} we show the dependence of the ratio
$R=\Gamma/3H$ and the tensor-to-scalar ratio $r$ on the
primordial tilt $n_s$ for the special case in which we fix $m=1$,
i.e., $\Gamma\propto T$, in the warm weak dissipative regime. In
both panels we have used three different values of the parameter
$C_\phi$. The upper panel shows the evolution of the rate
$R=\Gamma/3H$ during the inflationary scenario and we verify that
the rate $R < 1$. In the lower panel, we show the two-dimensional
marginalized constraints (68$\%$ and 95$\%$ C.L.) on
the inflationary parameters $r$ and $n_s$, defined as $k_0 = 0.002
Mpc^{-1}$, derived with the nine-year WMAP data with extended CMB (eCMB)
(green) and eCMB+BAO+$H_0$ (red); see Ref.\cite{astro202}. In
order to write down values for the ratio $R=\Gamma/3H$, $r$, and
$n_s$ for the case $m=1$, we utilize Eqs.
(\ref{awr}),(\ref{Ratio-ns}) and (\ref{Rk11}), where
$C_\gamma=70$, $f=0.30$, and $\kappa=1$. From the upper panel we
noted that the value $C_\phi<10^{-6}$ is well supported by the
weak regime ($R=\Gamma/3H< 1$). It is interesting to note that in
this case we have obtained an upper bound  for the parameter
$C_\phi$. From the lower panel we note that the value of the
parameter $C_\phi>10^{-9}$ is well supported by the confidence
levels from the nine-year WMAP data.

In Fig.\ref{fig2} we show the tensor-to-scalar
ratio $r$ versus the spectral index $n_s$ in the warm weak
dissipative regime for the case $m=-1$. In the upper panel we show
the two-dimensional marginalized constraints (68$\%$ and 95$\%$
C.L.) on inflationary parameters $r$ and $n_s$ from nine year WMAP.
In the lower panel, we shows the two-dimensional marginalized
constraints (68$\%$ and 95$\%$ C.L.) from Planck in conjunction
with Planck+WP Planck CMB temperature likelihood supplemented by
the WMAP large-scale polarization likelihood (grey),
Planck+WP+highL (red), and Planck+WP+BAO (blue) \cite{Planck}.
In both panels we have used three different values of the
parameter $C_\phi$. We note that the Planck data places stronger
limits on the tensor-to-scalar ratio $r$ versus $n_s$
compared with the nine-year WMAP data. From the lower panel in
which $m=-1$, we note that for the value $C_\phi>10^{-25}$ the
model is well supported by the Planck data. As before,  we note
that the value $C_\phi<10^{-20}$ is well supported from the
condition $R<1$ ( figure not shown).

Also, in particular for the value $m=0$ in this regime, we note
that for the value of the parameter $C_\phi>10^{-15}$ the model
is well supported by the Planck data. Also,  we noted that the
value $C_\phi<10^{-11}$ is well supported from the condition $R<1$
(figure not shown). In this form, for the value $m=0$ we get
$10^{-15}<C_\phi<10^{-11}$. We note that when we decrease the value
of the parameter $m$ the values of the parameters $C_\phi$ also
decrease.

%>From Eqs.(\ref{A}) and (\ref{Rk11}), we observed that for the
%special case in which $C_\phi=10^6$  and $f = \frac{1}{2}$, the
%curve $r = r(n_s)$ for WMAP 5-year enters the 95$\%$ confidence
%region for $r\simeq 0.26$, which corresponds to the number of
%e-folds, $N \simeq 146$. For $r \simeq 0.20$ corresponds to $N
%\simeq 140$, in this way the model is viable for large values of
%the number of e-folds.

\subsection{ Warm-Logamediate inflation\label{subsection2}}

Assuming that the system evolves according to the weak
dissipative regime and the scale factor is given by
Eq.(\ref{at1}), then from Eqs.(\ref{inf3}) and (\ref{at1}), we
find the solution of the scalar field as a function of the
cosmological time
\begin{equation}
\phi(t)=\sqrt{\frac{2\,A\,\lambda}{\kappa}}\left[\frac{2}{1+\lambda}\right](\ln
t)^{\frac{1+\lambda}{2}},\label{wrr1}
\end{equation}
where as before without loss of generality the integration constant
$\phi_0=0$. The Hubble parameter $H$ as a function of the inflaton
field  becomes $
H(\phi)=(A\,\lambda)B^{\lambda-1}\phi^{\gamma(\lambda-1)}\exp[-B\,\phi^{\gamma}],
$ where the constants $\gamma$ and $B$ are given by  $
\gamma=\frac{2}{\lambda+1}$ and $B\equiv\left[\frac{1}{\gamma}
\sqrt{\frac{\kappa}{2A\lambda}}\right]^{\gamma}$.

%From Eq.(\ref{pot}) we obtain that the scalar potential $V(\phi)$
%is given by
%\begin{equation}
%V(\phi)=C\phi^{-\beta_2}+D\phi^{-\beta_3},\label{pot1}
%\end{equation}
%where
%$$
%C=\frac{3\,f^2\,A^2}{\kappa}\,\left[
%\frac{f\,\kappa}{8\,A\,(1-f)}\right]^{2(f-1)/f},\,\;\;\beta_2=\frac{4(1-f)}{f},
%$$
%$$
%D=\frac{f\,A\,(f-1)}{\kappa}\,\left[
%\frac{f\,\kappa}{8\,A\,(1-f)}\right]^{(f-2)/f},\;\;\;\mbox{and}
%\,\,\;\beta_3=\frac{2\,(2-f)}{f}.
%$$
From Eq.(\ref{pot}) the effective potential with this scale factor
becomes
\begin{equation}
V(\phi)=V_{0}\phi^{\alpha}\exp[-\beta\,\phi^{\gamma}],\label{pot11}
\end{equation}
where $ V_{0}=\frac{3}{\kappa}(A\lambda)^{2}B^{2(\lambda-1)},
\,\;\;\alpha=2\gamma(\lambda-1)$, and $\beta=2B$.

Note that the potential given by Eq.(\ref{pot11})coincides with
the scalar potential obtained in Ref.\cite{R12}. Also, we note
that the scalar field $\phi$, the Hubble parameter $H$, and the
potential $V(\phi)$  become independent of the parameters
$C_\phi$ and $C_\gamma$ in the weak regime.

Considering Eq.(\ref{G1}), the dissipation coefficient $\Gamma$ in
terms of the scalar field is
\begin{equation}
\Gamma(\phi)=\, C_{\phi}^{{4 \over 4-m}}\left[\frac{1}{2\kappa\,
C_\gamma}\right]^{{m \over 4-m}}\phi^{{4(1-m) \over 4-m}}
 \exp[{-mB \over 4-m}\,\phi^{\gamma}].
\end{equation}

For this scale factor, the dimensionless slow-roll parameter
$\varepsilon$ and $\eta$ are given by $
\varepsilon=-\frac{\dot{H}}{H^2}=(A\lambda)^{-1}B^{-(\lambda-1)}\phi^{-\gamma(\lambda-1)}
$ and $ \eta=-\frac{\ddot{H}}{H \dot{H}}=(A\lambda
\,B^{\lambda})^{-1}\phi^{-\gamma\lambda}\left[2B\phi^{\gamma}-(\lambda-1)\right]\,.
$ Analogously as before, the condition for inflation to occur
is $\varepsilon<$1, and this is satisfied
when $\phi>[A\lambda
B^{(\lambda-1)}]^{\frac{-1}{\gamma(\lambda-1)}}$. Also
considering that inflation begins at the earliest possible stage,
where $\varepsilon=1$, the scalar field $\phi_1$ becomes $
\phi_{1}=[A\lambda
B^{(\lambda-1)}]^{\frac{-1}{\gamma(\lambda-1)}}\;. $

From Eq.(\ref{wrr1}), the number of $e$-folds $N$ between two
different values of the scalar field $\phi_1$ and $\phi_2$ is
\begin{equation}
N=\int_{t_1}^{t_{2}}\,H\,dt=A\,\left[(\ln t_{2})^{\lambda}-(\ln
t_{1})^{\lambda}\right]=A
B^{\lambda}\,\left(\phi^{\gamma\lambda}_{2}-\phi^{\gamma\lambda}_{1}\right).\label{NN1}
\end{equation}

On the other hand, the density perturbation could be written from
Eqs.(\ref{pd}) and (\ref{wrr1}) in terms of the scalar field as
\begin{equation}
{\cal{P}_{\cal{R}}}=
\frac{\kappa}{2}\left(\frac{C_{\phi}}{2\kappa C_{\gamma}}\right)^{{1 \over 4-m}}(A\lambda)^{2}B^{2(\lambda-1)}
\;\phi^{\alpha+{1-m \over 4-m}}\exp[-{5-m \over 4-m}B\,\phi^{\gamma}], \label{pd2}
\end{equation}
and the power spectrum in terms of the number of $e$-folds $N$ is

\begin{equation}
{\cal{P}_{\cal{R}}}=\beta_1\;
\left[\frac{N}{A}+(A\,\lambda)^{\frac{-\lambda}{\lambda-1}}\right]^{\frac{2(\lambda-1)}{\lambda}+
{(\lambda+1)(1-m) \over 2\lambda (4-m)}}
\exp
\left[-{5-m \over 4-m}\left[\frac{N}{A}+(A\,\lambda)^{\frac{-\lambda}{\lambda-1}}\right]^{\frac{1}{\lambda}}\right], \label{pd2}
\end{equation}
where the constant $\beta_1$ is given by $ \beta_1=
\frac{\kappa}{2}\left(\frac{C_{\phi}}{2\kappa
C_{\gamma}}\right)^{{1 \over 4-m}}(A\lambda)^{2}
B^{-{(\lambda+1)(1-m) \over 2(4-m)}}. $

The scalar spectral index $n_s$, from Eqs. (\ref{wrr1}) and
(\ref{pd2}), is given by
\begin{equation}
n_s=1-\frac{(5-m)B^{-(\lambda-1)}}{A\,\lambda(4-m)}\phi^{-\gamma(\lambda-1)}
+\left[\frac{2(\lambda-1)}{A\lambda}+ {(\lambda+1)(1-m) \over
2A\lambda
(4-m)}\right]B^{-\lambda}\phi^{-\gamma\lambda}.\label{nsss1}
\end{equation}

As before, we note that the scalar index  can be re-expressed in
terms of the number  $N$ as
\begin{equation}
n_s=1-\frac{(5-m)}{A\,\lambda(4-m)}
\left[\frac{N}{A}+(A\,\lambda)^{\frac{-\lambda}{\lambda-1}}\right]^{-\frac{\lambda-1}{\lambda}}
+\left[\frac{2(\lambda-1)}{A\lambda}+ {(\lambda+1)(1-m) \over
2A\lambda (4-m)}\right]
\left[\frac{N}{A}+(A\,\lambda)^{\frac{-\lambda}{\lambda-1}}\right]^{-1},\label{nslogadeb}
\end{equation}
where we have used  Eq.(\ref{NN1}).

For the  the tensor-to-scalar ratio $r$, from Eq.(\ref{pd2}) we have

\begin{equation}
r \simeq \frac{4}{\pi^2}\left(\frac{2\kappa
C_{\gamma}}{C_{\phi}}\right)^{{1 \over 4-m}} \;\phi^{-{1-m \over
4-m}}\exp\left[\left({5-m \over
4-m}-2\right)B\,\phi^{\gamma}\right]. \label{Rk}\end{equation}

Also, the tensor-to-scalar ratio as function of the number of
$e$-foldings $N$ becomes

\begin{equation}
r \simeq \beta_2\;
\left[\frac{N}{A}+(A\,\lambda)^{\frac{-\lambda}{\lambda-1}}\right]^{-{(\lambda+1)(1-m)
\over 2\lambda (4-m)}} \exp \left[\left({5-m \over
4-m}-2\right)\left[\frac{N}{A}+(A\,\lambda)^{\frac{-\lambda}{\lambda-1}}\right]^{\frac{1}{\lambda}}\right].
\label{Rk112}\end{equation} where the constant $ \beta_2=
\frac{4}{\pi^2}\left(\frac{2\kappa
C_{\gamma}}{C_{\phi}}\right)^{{1 \over 4-m}} B^{{(\lambda+1)(1-m)
\over 2(4-m)}} $.
\begin{figure}[th]

{\hspace{0.3cm}
\includegraphics[width=2.5in,angle=0,clip=true]{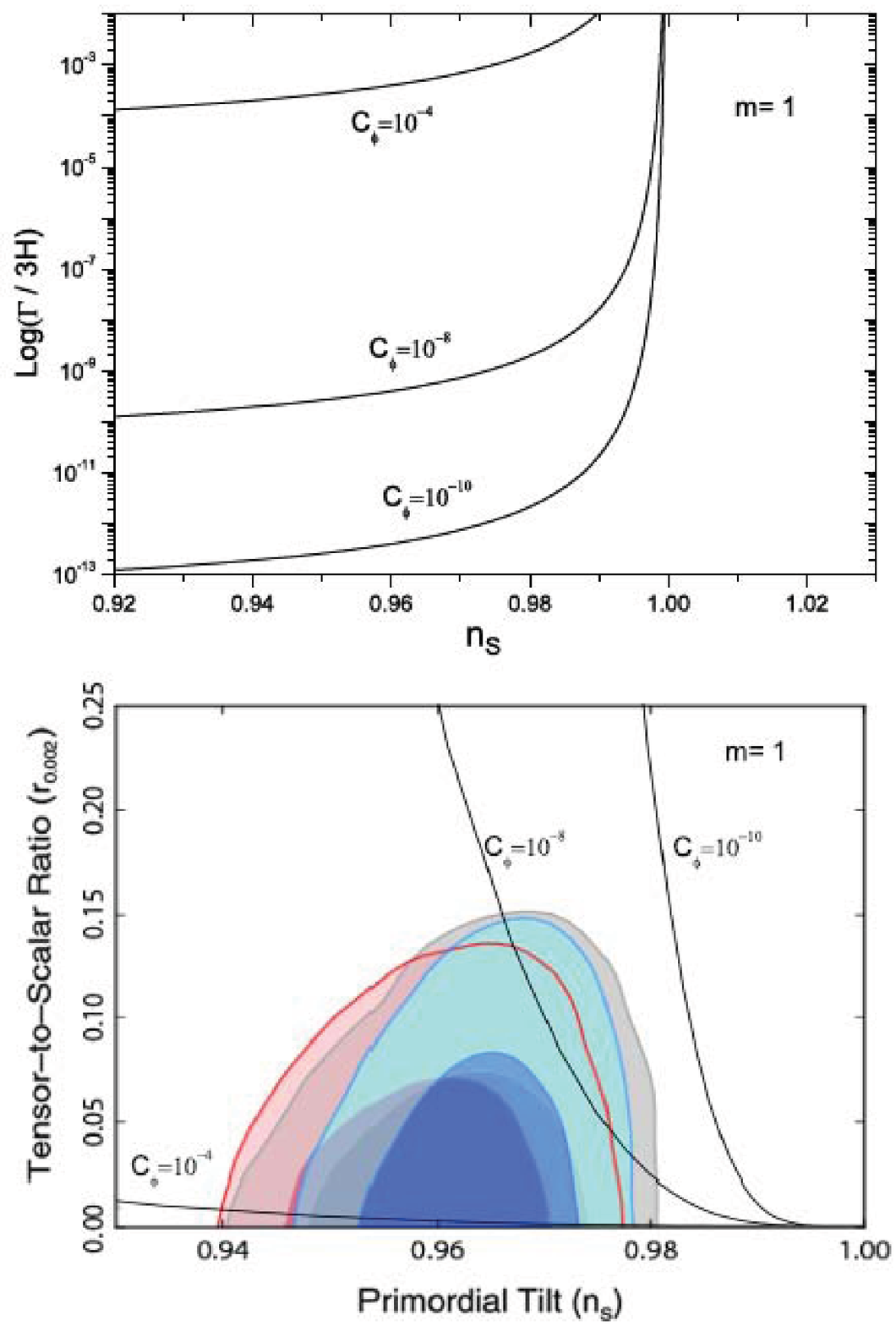}}
\caption{The evolution of the ratio
$R=\Gamma/3H$ versus the primordial tilt $n_s$ (upper panel) and the
evolution of the tensor-to-scalar ratio r versus $
n_s$ (lower panel) in the warm logamediate weak dissipative regime for the
case $m=1$, i.e., $\Gamma\propto T$. In both panels we use three
different values of the parameter $C_\phi$,
$\kappa=1$, and $C_\gamma=70$. In the lower panel we show the
two-dimensional marginalized constraints (68$\%$ and 95$\%$
C.L.) on the inflationary parameters $r$ and $n_s$,
derived from Planck \cite{Planck} \label{fig3}}.
\end{figure}

\begin{figure}[th]
\includegraphics[width=2.5in,angle=0,clip=true]{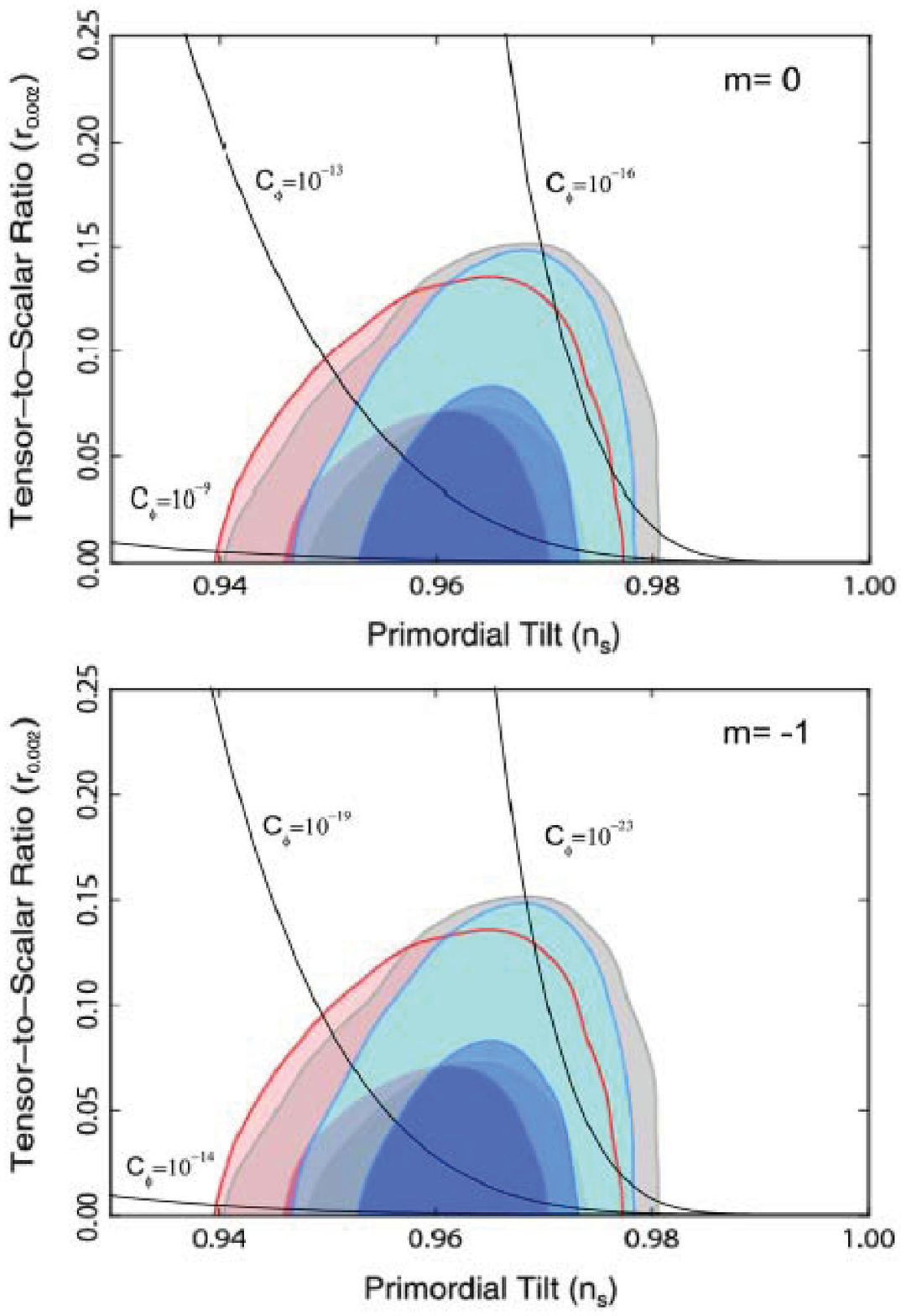}
\caption{The upper and lower panels show the evolution of the of
the tensor-to-scalar ratio r versus $ n_s$, in the warm
logamediate weak dissipative regime for the cases $m=0$ and
$m=-1$, respectively. In both panels we use three different
values of the parameter $C_\phi$,
$\kappa=1$ and $C_\gamma=70$. In both panels we show
the two-dimensional marginalized constraints (68$\%$ and 95$\%$
C.L.) on the inflationary parameters $r$ and $n_s$
derived from Planck \cite{Planck} \label{fig4}}.
\end{figure}

In Fig.\ref{fig3} we show the dependence of the ratio
$R=\Gamma/3H$ and the tensor-to-scalar ratio $r$ on the
primordial tilt $n_s$ for the special case in which we fixe $m=1$
in the warm logamediate weak dissipative regime. In both panels we
use three different values of the parameter $C_\phi$. In the
upper panel we show the decay of the ratio $R=\Gamma/3H$ during the
inflationary scenario and we also verify that the rate $R < 1$. In the
lower panel we show the two-dimensional marginalized
constraints (68$\%$ and 95$\%$ C.L.) on the inflationary parameters $r$
and $n_s$ derived from Planck.  In order to write down values for
the ratio $R=\Gamma/3H$, $r$, and $n_s$ for the case $m=1$, we
numerically  utilize Eqs. (\ref{nsss1}) and (\ref{Rk}), where
$C_\gamma=70$ and $\kappa=1$. Also, we numerically resolve
Eqs.(\ref{pd2}) and (\ref{nslogadeb}) and  we find that
$A=6.65\times10^{-5}$ and $\lambda= 5.03$ for the value of
$C_\phi=10^{-4}$, in which $n_s=0.96$, $N=60$, and
${\cal{P}_{\cal{R}}}=2.43\times10^{-9}$ .
Analogously, for $C_\phi=10^{-8}$, $A=8.43\times
10^{-4}$ and $\lambda=4.36$, and for $C_\phi=10^{-10}$,
 $A=2.88\times 10^{-3}$ and $\lambda=4.02$. From the
upper panel we note that the value $C_\phi<10^{-4}$ is well
supported by the weak regime ($R=\Gamma/3H< 1$). It is interesting
to note that in this case we have obtained an upper bound  for the
parameter $C_\phi$. From the lower panel we note that the value
of the parameter $C_\phi>10^{-10}$ is well supported by the
confidence levels from the Planck data.

In Fig.\ref{fig4} we show the dependence of the tensor-to-scalar
ratio on the spectral index for the weak regime in warm
logamediate inflation, where as before we use three different
values of the parameter $C_\phi$. In the upper panel we use $m=0$
and in the lower panel $m=-1$.
%From Ref.\cite{astro}, two-dimensional marginalized constraints
%(68$\%$ and 95$\%$ confidence levels) on the inflationary
%parameters $r$  and $n_s$, defined at $k_0$ = 0.002 Mpc$^{-1}$.
\begin{table}
\begin{tabular}
[c]{||l||l||l||l||}\hline\hline Regime & Scale factor &
$\Gamma=C_{\phi}\frac{T^{m}}{\phi^{m-1}}$ & \ Constraint on
$C_{\phi}$\\\hline\hline Weak & Intermediate $a(t)=e^{At^{f}}$ &
\begin{tabular}
[c]{c}%
$m=1$\\
$m=0$\\
$m=-1$%
\end{tabular}
&
\begin{tabular}
[c]{c}%
$10^{-9}<C_{\phi}<10^{-6}$\\
$10^{-15}<C_{\phi}<10^{-11}$\\
$10^{-25}<C_{\phi}<10^{-20}$%
\end{tabular}
\\\hline\hline
$\Gamma<3H$ & Logamediate $a(t)=e^{A(\ln t)^{\lambda}}$ &
\multicolumn{1}{||c||}{%
\begin{tabular}
[c]{c}%
$m=1$\\
$m=0$\\
$\ m=-1$%
\end{tabular}
} &
\begin{tabular}
[c]{c}%
$10^{-10}<C_{\phi}<10^{-4}$\\
$10^{-17}<C_{\phi}<10^{-6}$\\
$\ 10^{-24}<C_{\phi}<10^{-10}$%
\end{tabular}
\\\hline\hline
\end{tabular}
\caption{Results for the constraints on the parameter $C_\phi$ in
the weak regime. }
\end{table}

The Planck data places stronger limits on the tensor-to-scalar
ratio $r$. In order to write down values that relate $n_s$ and
$r$, we numerically solve Eqs.(\ref{nsss1}) and (\ref{Rk}). Also,
in both panels we have used the values $C_\gamma=70$ and
$\kappa=1$. Here, for the rate $R=\Gamma/3H$, $r$, and $n_s$ for
the case $m=0$, we numerically utilize Eqs. (\ref{nsss1}) and
(\ref{Rk}), where $C_\gamma=70$ and $\kappa=1$. Also, we
numerically resolve Eqs.(\ref{pd2}) and (\ref{nslogadeb}), and  we
find that $A=1.79\times10^{-4}$ and $\lambda= 3.78$ for the value
of $C_\phi=10^{-9}$, where $n_s=0.96$, $N=60$, and
${\cal{P}_{\cal{R}}}=2.43\times10^{-9}$.
Analogously, $C_\phi=10^{-13}$ corresponds to $A=1.16\times
10^{-3}$ and $\lambda=4.15$, and $C_\phi=10^{-16}$
corresponds to $A=4.53\times 10^{-3}$ and $\lambda=4.27$. From the
upper panel in which $m=0$, we note that for the value of the
parameter $C_\phi>10^{-17}$ the model is well supported by the
data in the warm logamediate weak regime. Also, we note that  for
$m=0$ the value $C_\phi<10^{-6}$ is well supported by the
condition $R=\Gamma/3H<1$ (not shown). In this form, for $m=0$ the
constraint for $C_\phi$ is given by $10^{-17}<C_\phi<10^{-6}$ for
the weak regime in logamediate inflation. As before, for the case
$m=-1$ we find that $A=3.15\times10^{-4}$ and $\lambda= 4.42$ for
$C_\phi=10^{-14}$, where $n_s=0.96$, $N=60$ and
${\cal{P}_{\cal{R}}}=2.43\times10^{-9}$.
Analogously, $C_\phi=10^{-19}$ corresponds to $A=1.99\times
10^{-3}$ and $\lambda=3.93$ and $C_\phi=10^{-23}$
corresponds to $A=8.33\times 10^{-3}$ and $\lambda=3.55$. Also,
from the lower panel in which $m=-1$ we note that for
$C_\phi>10^{-24}$ the model is well supported by Planck. As
before, we note that for $m=-1$ the value $C_\phi<10^{-10}$ is
well supported by the condition $R=\Gamma/3H<1$ (not shown) and
the constraint for $C_\phi$ is $10^{-24}<C_\phi<10^{-10}$. We
observe that when we decrease the values of the parameter $m$
the value of  $C_\phi$ decreases as well.

Table I  indicates the constraints of the parameter $C_\phi$ in
the weak regime  and different choices of the parameter $m$, for
a general form for the dissipative coefficient
$\Gamma(T,\phi)=C_\phi\,T^{m}/\phi^{m-1}$, in the context of warm
intermediate and logamediate inflationary universe models.

%>From Eqs.(\ref{A}) and (\ref{Rk11}), we observed that for the
%special case in which $C_\phi=10^6$  and $f = \frac{1}{2}$, the
%curve $r = r(n_s)$ for WMAP 5-year enters the 95$\%$ confidence
%region for $r\simeq 0.26$, which corresponds to the number of
%e-folds, $N \simeq 146$. For $r \simeq 0.20$ corresponds to $N
%\simeq 140$, in this way the model is viable for large values of
%the number of e-folds.

\section{ The Strong dissipative regime $\Gamma>3H$\label{section4}}

\subsection{ Warm-Intermediate inflation\label{subsection1}}

We now analyze the case of the strong dissipative regime
($R=\Gamma/3H>1$), together with  the scale factor given by
Eq.(\ref{at}), i.e., intermediate inflation. From Eqs.(\ref{inf3})
and (\ref{G1}) we get $
\dot{\phi}\phi^{\frac{1-m}{2}}=k_{6}t^{\beta_{6}}, $ where
$k_{6}=\left[  \frac{2}{\kappa\alpha_{m}}(\left[ 3(1-f)\right]
^{4-m}(Af)^{8-m})^{\frac{1}{4}}\right]  ^{\frac{1}{2}}$ and $\beta_{6}%
=\frac{1}{8}\left[  f(8-m)+2m-12\right]  $. In this way, the
solution of the scalar field $\phi(t)$ is given by
\begin{equation}
\phi(t)-\phi_0=k_{7}\,t^{\frac{f(8-m)+2m-4}{4(3-m)}}, \label{phsr}%
\end{equation}
where the constant $k_{7}=\left[  \frac{4k_{6}(3-m)}{f(8-m)+2m-4}\right]  ^{\frac{2}{3-m}}%
$ and $\phi(t=0)=\phi_0$  is an integration constant. As before,
without loss of generality, we consider
 $\phi_{0}=0$. The Hubble parameter
as a function of the inflaton field is given by  $H(\phi) =Af\left(
\frac{\phi}{k_{7}}\right) ^{\frac{-4(1-f)(3-m)} {f(8-m)+2m-4}}.$

From Eq.(\ref{pot}) the scalar potential as a function of the scalar
field is
\begin{equation}
V(\phi)=\frac{3\left(  Af\right)  ^{2}}{\kappa}k_{7}^{\frac{8(1-f)(3-m)}%
{f(8-m)+2m-4}}\phi^{\frac{-8(1-f)(3-m)}{f(8-m)+2m-4}}. \label{potsr}%
\end{equation}
Here we note that in the case of the strong regime  the scalar
field $\phi$, the Hubble parameter $H$, and the potential $V(\phi)$
now depend on the parameters $C_{\phi}$ and $C_{\gamma}$.

The dissipation coefficient $\Gamma$ in terms of the scalar field
considering  Eq.(\ref{G1}) is given by
\begin{equation}
\Gamma(\phi)=k_{8}\phi^{\beta_{7}}, \label{gammaphsr}%
\end{equation}
where the constants $k_8$ and $\beta_7$ are defined as $k_{8}=\alpha_{m}3^{\frac{m}{4}}\left[  Af(1-f)\right]  ^{\frac{m}{4}%
}k_{7}^{\frac{m(2-f)(3-m)}{f(8-m)+2m-4}}$ and $\beta_{7}=\frac{f(8-6m)-4}%
{f(8-m)+2m-4}$.

For this regime, the dimensionless slow-roll parameter
$\varepsilon$ becomes $
\varepsilon=-\frac{\dot{H}}{H^{2}}=\frac{(1-f)}{Af}\left(  \frac{\phi}{k_{7}%
}\right)  ^{-\beta_{8}},%
$
%and the other slow-roll parameter $\eta$, becomes%
%\begin{equation}
%\eta=-\frac{\ddot{H}}{H\dot{H}}=\frac{(2-f)}{Af}\left(  \frac{\phi}{k_{7}%
%}\right)  ^{-\beta_{8}},\label{etsr}%
%\end{equation}
where the constant $\beta_{8}=\frac{4f(3-m)}{f(8-m)+2m-4}$. As
before, the condition
for inflation to occur  is only satisfied when $\phi>k_{7}\left(  \frac{1-f}%
{Af}\right)  ^{\frac{1}{\beta_{8}}}.$ Also,  considering that
inflation begins at the earliest possible stage
(where $\varepsilon=1$), we get $
\phi_{1}=k_{7}\left(  \frac{1-f}{Af}\right)  ^{\frac{1}{\beta_{8}}%
}. $

In this regime, the number of $e$-folds $N$ between two different
 values of the scalar field
$\phi_{1}$ and $\phi_{2}$ from Eq.(\ref{phsr}) is
\begin{equation}
N=\int_{t_{1}}^{t_{2}}\,H\,dt=A\,\left(  t_{2}^{f}-t_{1}^{f}\right)
=Ak_{7}^{-\beta_{8}}\left[  \phi_{2}^{\beta_{8}}-\phi_{1}^{\beta_{8}}\right]
.\label{Nsr}%
\end{equation}

On the other hand, as the  scalar perturbations, ${\mathcal{P}%
_{\mathcal{R}}}^{1/2}\propto\frac{H}{\dot{\phi}}\,\delta\phi$, where
now $\delta\phi ^{2}$
%in the case of high dissipation, following Taylor and
%Berera\bigskip\cite{TAYBER}, or
in the strong dissipation regime   is given by, $\delta\phi
^{2}\simeq\frac{k_{F}T}{2\pi^{2}}$; see Ref.\cite{taylorberera}.
Here $k_{F}$ is the wave-number and it is given by
$k_{F}=\sqrt{\Gamma H}=H\sqrt{3R}$. In this form, by combining the
Eqs.(\ref{inf3}), (\ref{rh-1}), and (\ref{G1}), the expression for
the spectrum of the scalar perturbation can be written as
\begin{equation}
P_{\mathcal{R}}\simeq\frac{H^{\frac{5}{2}}\Gamma^{\frac{1}{2}}T}{2\pi^{2}%
\dot{\phi}^{2}}=\frac{\kappa C_{\phi}^{\frac{3}{2}}3^{\frac{3m-6}{8}}}%
{4\pi^{2}\left(  2\kappa C_{\gamma}\right)  ^{\frac{3m+2}{8}}}H^{\frac{3}{2}%
}\left(  -\dot{H}\right)  ^{\frac{3m-6}{8}}\phi^{\frac{3(1-m)}{2}%
}.\label{Prsr}%
\end{equation}
By using Eqs. (\ref{phsr}) and (\ref{Prsr}), the power spectrum
in terms of the scalar field can be written as
\begin{equation}
P_{\mathcal{R}}=k_{9}\phi^{-\beta_{9}},\label{Prphsr}%
\end{equation}
where $k_{9}=\frac{\kappa C_{\phi}^{\frac{3}{2}}3^{\frac{3m-6}{8}}\left(
Af\right)  ^{\frac{3m+6}{8}}\left(  1-f\right)  ^{\frac{3m-6}{8}}}{4\pi
^{2}\left(  2\kappa C_{\gamma}\right)  ^{\frac{3m+2}{8}}}k_{7}^{\frac{3}%
{2}(1-m)+\beta_{9}}$ and
$\beta_{9}=\frac{3(2+f(4m-7))}{f(8-m)+2m-4}.$

\bigskip

The scalar spectral index $n_{s}=n_{s}(\phi)$ from Eqs.
(\ref{phsr}) and (\ref{Prphsr}) is
\begin{equation}
n_{s}=1-\frac{3\left[  2+f(4m-7)\right]  }{4Af(3-m)}k_{7}^{\beta_{8}}%
\phi^{-\beta_{8}}.\label{nsphsr}%
\end{equation}
Note that the spectral index, can also be re-expressed in terms of
the number of $e$-foldings. Combining Eqs.(\ref{Nsr}) and
(\ref{nsphsr}), the spectral index becomes

\begin{equation}
n_{s}=1-\frac{3(2+f(4m-7))}{4(3-m)(1+f(N-1))}.\label{nsMsr}%
\end{equation}
As before, here the value of $f$ in terms of $m$, $n_{s}$, and
$N$ from Eq.(\ref{nsMsr}) is given by $
f=\frac{4(1-n_{s})(3-m)-6}{3(4m-7)-4(3-m)(1-n_{s})(N-1)}. \label{ffsr}%
$ In particular, for the values $m=1$, $n_s=0.96$, and $N=60$ we
obtain $f\simeq 0.21$.
Analogously, $m=0$ corresponds to $f\simeq 0.11$, and
$m=-1$ corresponds to $f\simeq0.08$.

Analogously to the case of the intermediate weak regime,  we can
obtain an analytic expression for the parameter $A$ in terms of the parameters $C_{\phi}$, $C_{\gamma}%
$, ${\mathcal{P}_{\mathcal{R}}}$, $m$, $n_{s}$, and $N$. Considering
Eqs.(\ref{Nsr}) and (\ref{Prphsr}), we get
\begin{equation}
A=\frac{1}{f}\left(  \frac{{\mathcal{P}_{\mathcal{R}}}}{k_{10}}\right)
^{\frac{2f(3-m)}{3}}\left[  f(N-1)+1\right]  ^{\frac{2+f(4m-7)}{2}%
},\label{Aff}%
\end{equation}
where the constant $k_{10}=\frac{\kappa
C_{\phi}^{\frac{3}{2}}3^{\frac{3m-6}{8}}\left( Af\right)
^{\frac{3m+6}{8}}\left(  1-f\right)  ^{\frac{3m-6}{8}}}{4\pi
^{2}\left(  2\kappa C_{\gamma}\right)  ^{\frac{3m+2}{8}}}\left[
\frac {\beta_{8}}{f}\left(  \frac{2}{\kappa\alpha_{m}}\left[
3(1-f)\right] ^{\frac{4-m}{4}}\right)  ^{\frac{1}{2}}\right]
^{\frac{3(1-m)}{3-m}}.$ In particular, $m=1$, where
 $f\simeq 0.21$, $n_s=0.96$, $C_\gamma=70$,
${\mathcal{P}_{\mathcal{R}}}=2.4\times 10^{-9}$, and $N=60$, we
obtain that $A\simeq1.76$ when
$C_\phi=2\times 10^{-1}$, $A=1.21$ when $C_\phi=5\times10^{-1}$,
and $A=0.92$ when $C_\phi=1$ .  Analogously, for
$m=0$, $A=10.33$ when $C_\phi=10^{-5}$, $A=3.68$ when $C_\phi=10^{-3}$, and $A=1.54$ when $C_\phi=5\times10^{-2}$. Finally, for
$m=-1$, $A=22.40$ when $C_\phi=10^{-9}$, $A=5.55$ when $C_\phi=10^{-5}$, $A=0.97$ when $C_\phi=1$.

\bigskip

Also, we can find an expression for the rate $R=\Gamma/3H$ in
terms of the scalar spectral index. By using Eqs.(\ref{gammaphsr})
and (\ref{nsphsr}) the ratio $R$ has the following dependence on
$n_{s}$:
\begin{equation}
R(n_{s})=\frac{k_{8}k_{7}^{\beta_{7}}}{3Af}\left(  \frac{3\left[
2+f(4m-7)\right]  }{4Af(3-m)(1-n_{s})}\right)  ^{-\frac{\left[
4(m-2)+2f(m+2)\right]  }{4f(3-m)}}. \label{Rns}%
\end{equation}
\begin{figure}[th]
\includegraphics[width=2.5in,angle=0,clip=true]{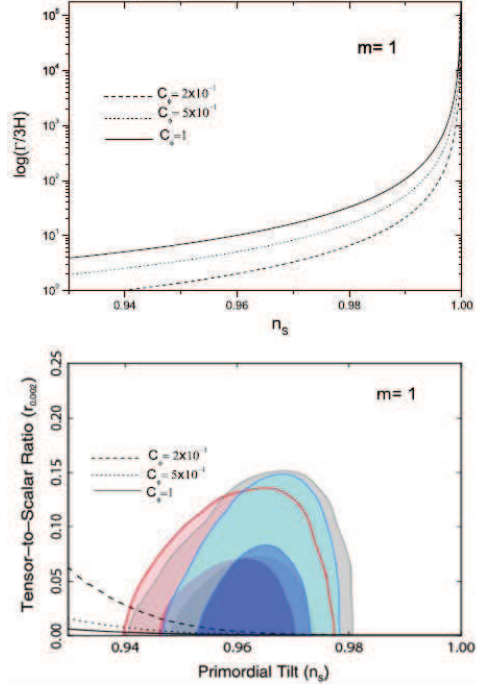}

{\hspace{0.3cm}}
\caption{The evolution of the ratio
$\log(\Gamma/3H)$ versus the primordial tilt $n_s$ (upper panel) and the
evolution of the tensor-to-scalar ratio r versus $
n_s$ (lower panel) in the warm intermediate strong dissipative regime for the
case $m=1$, i.e., $\Gamma\propto T$. In both panels we use three
different values of the parameter $C_\phi$, and
$\kappa=1$ and $C_\gamma=70$. In the lower panel we show the
two-dimensional marginalized constraints (68$\%$ and 95$\%$ C.L.) on
the inflationary parameters $r$ and $n_s$ derived from Planck
\cite{Planck} \label{fig5}}.
\end{figure}

\begin{figure}[th]
\includegraphics[width=2.5in,angle=0,clip=true]{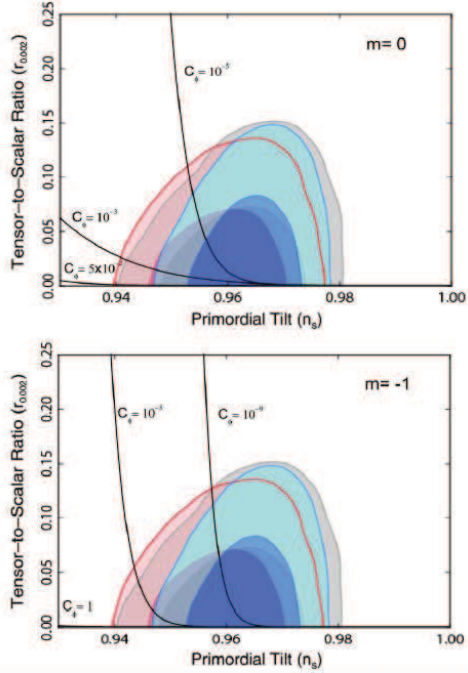}
\caption{The upper and lower panel show the evolution of
the tensor-to-scalar ratio r versus $ n_s$ in the warm
intermediate strong dissipative regime for the cases $m=0$ and
$m=-1$, respectively. As before, In both panels, we use three
different values of the parameter $C_\phi$, and
$\kappa=1$ and $C_\gamma=70$. Also, in both panels we show the
two-dimensional marginalized constraints (68$\%$ and 95$\%$
C.L.) on the inflationary parameters $r$ and $n_s$
derived from Planck; see Ref.\cite{Planck} \label{fig6}}.
\end{figure}

%\bigskip

On the other hand, for the intermediate strong dissipative regime,
the tensor-to-scalar ratio $r=r(\phi)$ from   Eqs.(\ref{phsr}) and
(\ref{Prphsr}) is
\begin{equation}
r(\phi)=\left(  \frac{{\mathcal{P}}_{g}}{P_{\mathcal{R}}}\right)
\simeq
\frac{2\kappa\left(  Af\right)  ^{2}k_{7}^{\beta_{10}}}{\pi^{2}k_{9}}%
\phi^{\beta_{11}}, \label{rphsr}%
\end{equation}
where the constants $\beta_{10}$ and $\beta_{11}$ are defined as $\beta_{10}=\frac{8(1-f)(3-m)}{f(8-m)+2m-4}$ and $\beta_{11}%
=\frac{3(f-6)+4m(2+f)}{f(8-m)+2m-4}$.

Also, the tensor-to-scalar ratio can be written in terms of the scalar
spectral
index $n_{s}$ as%
\begin{equation}
r(n_{s})=\frac{2\kappa\left(  Af\right)  ^{2}k_{7}^{\beta_{10}+\beta_{11}}%
}{\pi^{2}k_{9}}\left[  \frac{3\left[  2+f(4m-7)\right]  }{4Af(3-m)(1-n_{s}%
)}\right]  ^{\frac{3(f-6)+4m(2+f)}{4f(3-m)}}, \label{rnssr}%
\end{equation}
and also  the tensor-to-scalar ratio as a function of the number of
$e$-foldings $N$ is
\begin{equation}
r(N)=\frac{2\kappa(Af)^{2}k_{7}^{\beta_{10}+\beta_{11}}}{\pi^{2}k_{9}}\left[
\frac{f(N-1)+1}{Af}\right]  ^{\frac{3(f-6)+4m(2+f)}{4f(3-m)}}. \label{RNSR}%
\end{equation}

In Fig.\ref{fig5}, we show the dependence of the ratio
$\log(\Gamma/3H)$ and the tensor-scalar ratio $r$ on the
primordial tilt $n_s$ for the special case in which we fix $m=1$
in the warm intermediate strong dissipative regime.  As before, in
both panels we have used three different values of the parameter
$C_\phi$. In the upper panel we show the decay of the ratio
$R=\Gamma/3H$ during the inflationary scenario and  we verify that
the rate $R < 1$. In the lower panel we show the two-dimensional
marginalized constraints (68$\%$ and 95$\%$ C.L.) on the inflationary
parameters $r$ and $n_s$ derived from Planck\cite{Planck}.  In
order to write down values for the ratio $R=\Gamma/3H$, $r$, and
$n_s$ for the case $m=1$, we   consider Eqs. (\ref{Rns}) and
(\ref{rnssr}) (where $C_\gamma=70$,$\kappa=1$) and the values of
$f$ and $A$ for $m=1$ obtained from Eqs.(\ref{nsMsr}) and
(\ref{Aff}). From the upper panel we note that the value
$C_\phi>2\times 10^{-1}$ is well supported by the strong regime
($R=\Gamma/3H> 1$). From the lower panel we note that the value
$C_\phi<1$ is well supported by the confidence
levels from Planck. In this form, the range of the parameter
$C_\phi$ for the value $m=1$ is $2\times
10^{-1}<C_\phi<1$.

In Fig.\ref{fig6}, we show the dependence of the tensor-to-scalar
ratio $r$ on the primordial tilt $n_s$ for the cases $m=0$ and
$m=-1$ in the warm intermediate strong dissipative regime.  As
before, in both panels we have used three different values of the
parameter $C_\phi$. In both panels, we show the two-dimensional
marginalized constraints (68$\%$ and 95$\%$ C.L.) on the inflationary
parameters $r$ and $n_s$ derived from Planck. In order to write
down values for $r$ and $n_s$ for each value of $m$, we
 use Eq. (\ref{rnssr}) (where
$C_\gamma=70$,$\kappa=1$) and the values of $f$ and $A$ for each
value of $m$ and $C_\phi$ obtained from Eqs.(\ref{nsMsr}) and
(\ref{Aff}). From the upper panel in which $m=0$, we note that
the value $C_\phi<10^{-2}$ is well supported by the confidence
levels from the Planck data. Also, we observe that the value
$C_\phi>10^{-6}$ is well supported by the strong regime, i.e.,
$R=\Gamma/3H> 1$ (not shown). In this form, the range of the
parameter $C_\phi$ for $m=0$, is $10^{-6}<C_\phi<10^{-2}$.
 From the lower panel we note that the
value $C_\phi<1$ is well supported by the
confidence levels from the Planck data. Also, we note that the
value $C_\phi>10^{-11}$ is well supported by the strong
regime, i.e., $R=\Gamma/3H> 1$ (also not shown). In this way, the
range of the parameter $C_\phi$ for $m=-1$ is $10^{-11}<C_\phi<1$.

\subsection{ Warm logamediate inflation\label{subsection2}}

We now consider the case of the strong regime together with $a(t)$
given by Eq.(\ref{at1}). From Eq.(\ref{inf3}), the solution of
$\phi(t)$ is given by

\begin{equation}
\varphi(t)\equiv\left({2\over 3-m}\right)\phi(t)^{{3-m\over
2}}\,=\alpha_1\ \gamma_{m}[t],\label{wr12}
\end{equation}
where  the constant $\alpha_1$ is defined by $
\alpha_1={(4C_\gamma)^{m/8}\over C_{\phi}^{1/2}} \left({6\over
\kappa}\right)^{{4-m\over 8}} (A\,\lambda)^{{8-m\over 8}}\left({4
\over 2-m}\right)^{1+\frac{8-m}{8}(\lambda-1)}, $ and the function
$$
\gamma_{m}[t]\equiv\,\gamma\left[1+\frac{8-m}{8}(\lambda-1),\frac{2-m}{4}\ln
t\right]
$$
is the incomplete gamma function; see,e.g.,
Refs.\cite{Libro,Libro02}. The Hubble parameter  $H=H(\phi)$ is
given by the expression $
H(\phi)=A\,\lambda\,\left(\gamma_{m}^{-1}\left[\varphi/\alpha_1\right]\right)^{-1}
\left(\ln\gamma_{m}^{-1}\left[\varphi/\alpha_1\right]\right)^{\lambda-1},\label{HH2}
$ where $\gamma_{m}^{-1}\left[\varphi/\alpha_1\right]$ denotes the
inverse gamma function of $\gamma_{m}[t]$.
%\textbf{la funcion inversa}
%$\gamma_{\lambda}^{-1}\left[\frac{1}{\alpha_1}\;\ln\phi\right] $
%\textbf{ se calcula numericamente.}

Analogous to the case of the logamediate weak dissipative regime,
the scalar potential from Eq.(\ref{pot}) becomes
\begin{equation}
V(\varphi)=\frac{3}{\kappa}(A\,\lambda)^{2}
\left(\gamma_{m}^{-1}\left[\varphi/\alpha_1\right]\right)^{-2}
\left(\ln\gamma_{m}^{-1}\left[\varphi/\alpha_1\right]\right)^{2(\lambda-1)}
,\label{pot111}
\end{equation}
and  the dissipation coefficient $\Gamma=\Gamma(\phi)$ from
Eq.(\ref{G1}) is
\begin{equation}
\Gamma(\phi)= C_\phi\left[\frac{3}{2\kappa\,
C_\gamma}\right]^{m/4}(A\lambda)^{m/4}\phi^{1-m}\left(\gamma_{m}^{-1}\left[\varphi/\alpha_1\right]\right)^{-m/2}
\left(\ln\gamma_{m}^{-1}\left[\varphi/\alpha_1\right]\right)^{\frac{m}{4}(\lambda-1)}.\label{gg2}
\end{equation}

As before, the dimensionless slow-roll parameter $\varepsilon$ for
this regime becomes $
\varepsilon=-\frac{\dot{H}}{H^2}=(A\lambda)^{-1}\left(\ln\gamma_{m}^{-1}\left[\varphi/\alpha_1\right]\right)^{-(\lambda-1)},
$ and the slow-roll parameter  $ \eta=-\frac{\ddot{H}}{H
\dot{H}}=(A\lambda)^{-1}\left(\ln\gamma_{m}^{-1}\left[\varphi/\alpha_1\right]\right)^{-\lambda}
\left\{2\left(\ln\gamma_{m}^{-1}\left[\varphi/\alpha_1\right]\right)-(\lambda-1)\right\}.
$

%Again  $\eta$ reaches unity before $\varepsilon$ does. In this
%way, we may establish that the end of  inflation is governed by
%the condition $\eta=1$.
Again, following Ref.\cite{R12}, the condition $\varepsilon=1$ at
the beginning of inflation the scalar field gives
\begin{equation}
\varphi_{1}=\left({2\over 3-m}\right)\,\phi_{1}^{{3-m\over
2}}=\alpha_1\;\gamma_{m}\left[\exp[(A\lambda)^{\frac{-1}{\lambda-1}}]\right],
\label{al22}
\end{equation}

The number of $e$-folds $N$ in this regime from Eq.(\ref{wr12}) is
given by
\begin{equation}
N=\int_{t_1}^{t_{2}}\,H\,dt=A\,
\left\{\left(\ln\gamma_{m}^{-1}\left[\varphi_2/\alpha_1\right]\right)^{\lambda}
-\left(\ln\gamma_{m}^{-1}\left[\varphi_1/\alpha_1\right]\right)^{\lambda}\right\}.
\label{N22}
\end{equation}

%wave-number
%$k_F$ is defined at the point where the inequality
%$V_{,\,\phi\,\phi}< \Gamma H$, is satisfied \cite{Bere2}.
From Eqs.(\ref{Prsr}) and (\ref{wr12}) we obtain that
${\cal{P}_{\cal{R}}}$ in terms of the scalar field becomes
\begin{equation}
{\cal{P}_{\cal{R}}}=
\alpha_2\,\phi^{{3(1-m) \over 2}}\,\left(\gamma_{m}^{-1}\left[\varphi/\alpha_1\right]\right)^{-{3m \over 4}}
\left(\ln\gamma_{m}^{-1}\left[\varphi/\alpha_1\right]\right)^{\frac{3m+6}{8}(\lambda-1)}, \label{pd21}
\end{equation}
where the constant $\alpha_2$ is given by $
\alpha_2=\frac{\kappa}{4\pi^2}3^{{3m-6 \over 8}}C_{\phi}^{3/2}
\left(\frac{1}{2\kappa C_{\gamma}}\right)^{{3m+2 \over 8}}
(A\lambda)^{{3m+6 \over 8}}. $

Also, as before the scalar spectrum can be re-expressed in terms
of the number of e-folding $N$, as
\begin{equation}
{\cal{P}_{\cal{R}}}(N)= \alpha_2\,
\left[\frac{N}{A}+(A\,\lambda)^{\frac{-\lambda}{\lambda-1}}\right]^{
{(3m+6)(\lambda-1) \over 8\lambda }} \exp \left[-{3m \over
4}\left[\frac{N}{A}+(A\,\lambda)^{\frac{-\lambda}{\lambda-1}}\right]^{\frac{1}{\lambda}}\right]
\,F_m(N), \label{pd21}
\end{equation}
where the function
%$F_m(N)$ is defined as
$
F_m(N)=\left[\alpha_1\, \left({3-m \over
2}\right)\,\gamma_{m}(\exp
\left[\left[\frac{N}{A}+(A\,\lambda)^{\frac{-\lambda}{\lambda-1}}\right]^{\frac{1}{\lambda}}\right])
\right]^{{3(1-m)\over 3-m}}. $

%for $m = 3$

%$$
%F_3(N)=\exp\left[-3\alpha_1 \gamma_{3}[\exp
%\left[\left[\frac{N}{A}+(A\,\lambda)^{\frac{-\lambda}{\lambda-1}}\right]^{\frac{1}{\lambda}}\right]]
%\right]
%$$
Considering  Eqs.(\ref{wr12}) and (\ref{gg2}), the scalar spectral
index $n_s$ is given by
\begin{equation}
n_s=
1-\frac{3m}{4A\lambda}\left(\ln\gamma_{m}^{-1}\left[\varphi/\alpha_1\right]\right)^{-(\lambda-1)}
+\frac{(3m+6)(\lambda-1)}{8A\lambda}\left(\ln\gamma_{m}^{-1}\left[\varphi/\alpha_1\right]\right)^{-\lambda}
+K\,G_m(\phi).\label{nss}
\end{equation}
where the constant $K$ is defined as $ K={3(1-m) \over
2}C_{\phi}^{-1/2}\left(4 C_{\gamma}\right)^{{m \over 8}}
\left(\frac{6}{\kappa }\right)^{{4-m \over 8}} (A\lambda)^{-{m
\over 8}} $ and the function $ G_m(\phi)=\phi^{{m-3 \over
2}}\,\left(\gamma_{m}^{-1}\left[\varphi/\alpha_1\right]\right)^{{m-2
\over 4}}
\left(\ln\gamma_{m}^{-1}\left[\varphi/\alpha_1\right]\right)^{-\frac{m}{8}(\lambda-1)}
$.

Analogously, as before the scalar spectral index can be write in
terms of the number of $e$-folds. Considering Eqs. (\ref{al22}) and
(\ref{N22}), we obtain
\begin{equation}
n_s=
1-\frac{3m}{4A\,\lambda}\left[\frac{N}{A}+(A\,\lambda)^{\frac{-\lambda}{\lambda-1}}\right]^{-\frac{\lambda-1}{\lambda}}
+\frac{(3m+6)(\lambda-1)}{8A\,\lambda}\left[\frac{N}{A}+(A\,\lambda)^{\frac{-\lambda}{\lambda-1}}\right]^{-1}
+K\,J_m(N), \label{ns22}
\end{equation}
where $
J_m(N)=\left[\frac{N}{A}+(A\,\lambda)^{\frac{-\lambda}{\lambda-1}}\right]^{
{-m(\lambda-1) \over 8\lambda }} \exp \left[{m-2 \over
4}\left[\frac{N}{A}+(A\,\lambda)^{\frac{-\lambda}{\lambda-1}}\right]^{\frac{1}{\lambda}}\right]
\,j_m(N) $ and the function $j_m(N)$ is given by
$
j_m(N)=\left[\alpha_1 {3-m \over 2}\gamma_{m}[\exp
\left[\left[\frac{N}{A}+(A\,\lambda)^{\frac{-\lambda}{\lambda-1}}\right]^{\frac{1}{\lambda}}\right]]
\right]^{-1}.
$

For this regime and  scale factor,  we may write the tensor-to-scalar
ratio as
\begin{equation}
r =\,\alpha_3\,\phi^{{3(m-1) \over
2}}\,\left(\gamma_{m}^{-1}\left[\varphi/\alpha_1\right]\right)^{{3m-8
\over 4}}
\left(\ln\gamma_{m}^{-1}\left[\varphi/\alpha_1\right]\right)^{\frac{10-3m}{8}(\lambda-1)},
\label{Rk2}
\end{equation}
where the constant $
\alpha_3=\frac{2\kappa(A\lambda)^{2}}{\pi^{2}\alpha_{2}}$. Also,
we can write the tensor-to-scalar ratio as a function of the number of
$e$-foldings $N$ as
\begin{equation}
r =\alpha_3 \,
\left[\frac{N}{A}+(A\,\lambda)^{\frac{-\lambda}{\lambda-1}}\right]^{
{(10-3m)(\lambda-1) \over 8\lambda }} \exp \left[{3m-8 \over
4}\left[\frac{N}{A}+(A\,\lambda)^{\frac{-\lambda}{\lambda-1}}\right]^{\frac{1}{\lambda}}\right]
\,{1 \over F_m(N)}. \label{Rk22}
\end{equation}

%\begin{figure}[th]
%\includegraphics[width=6.0in,angle=0,clip=true]{figstrong.eps}
%\caption{Evolution of the tensor-scalar ratio $r$ versus the
%scalar spectrum index $n_s$ in the strong dissipative regime, for
%three different values of the parameter $C_\phi$. Here, we have
%used $f=1/2$, $\kappa=1$, $C_\gamma=70$  and
%${\cal{P}_{\cal{R}}}=2.4\times 10^{-9}$.
% \label{fig2}}
%\end{figure}

\begin{figure}[th]
\includegraphics[width=2.5in,angle=0,clip=true]{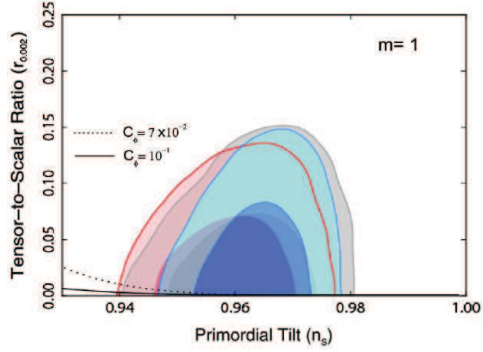}
\caption{The plot shows the evolution of the of the
tensor-to-scalar ratio r versus $ n_s$ in the warm
logamediate strong dissipative regime for the case $m=1$. Here
we use two different values of the parameter $C_\phi$, and $\kappa=1$ and $C_\gamma=70$. We show the
two-dimensional marginalized constraints (68$\%$ and 95$\%$
C.L.) on the inflationary parameters $r$ and $n_s$
derived from Planck data\cite{Planck} .\label{fig7}}
\end{figure}

In Fig.\ref{fig7}, we show the dependence of the tensor-to-scalar
ratio $r$ on the primordial tilt $n_s$ for the special case
in which we fix $m=1$ in the warm logamediate strong dissipative
regime. Here, we use two different values of the parameter
$C_\phi$, and we also show the two-dimensional marginalized
constraints (68$\%$ and 95$\%$ C.L.) on the inflationary parameters $r$
and $n_s$ derived from Planck\cite{Planck}.  In order to write
down values for the ratio  $r$ and $n_s$ for the case $m=1$, we
numerically utilize Eqs. (\ref{nss}) and (\ref{Rk2}), where
$C_\gamma=70$ and $\kappa=1$. Also, we numerically resolve
Eqs.(\ref{pd2}) and (\ref{nslogadeb}) and  we find that
$A=2.01\times10^{-3}$ and $\lambda= 3.71$ for the value of
$C_\phi=10^{-1}$, for which $n_s=0.96$, $N=60$, and
${\cal{P}_{\cal{R}}}=2.43\times10^{-9}$.
Analogously, $C_\phi=7\times10^{-2}$ corresponds to
$A=3.53\times 10^{-3}$ and $\lambda=3.56$. From the plot we note
that the value $C_\phi<10^{-1}$ is well supported by the
confidence levels from the Planck data, since for values of
$C_\phi>10^{-1}$ the ratio  $r\sim 0$. Also, we note
that the value of the parameter $C_\phi>7\times 10^{-2}$ is well
supported by the strong regime, i.e., $R=\Gamma/3H> 1$ (not shown).
In this way, the range for the parameter $C_\phi$ in the special
case in which $m=1$ is $7\times 10^{-2}<C_\phi<10^{-1}$.

Also, we note that for the cases in which $m=0$ and $m=-1$ the
models of the warm logamediate strong regime are disfavored from
the observational data, since the spectral index $n_s>1$, see Eq.(\ref{ns22})) and the models do not work.
\begin{table}
\begin{tabular}
[c]{||l||l||l||l||}\hline\hline Regime & Scale factor &
$\Gamma=C_{\phi}\frac{T^{m}}{\phi^{m-1}}$ & \ Constraint on
$C_{\phi}$\\\hline\hline Strong & Intermediate $a(t)=e^{At^{f}}$ &
\begin{tabular}
[c]{c}%
$m=1$\\
$m=0$\\
$\ m=-1$%
\end{tabular}
&
\begin{tabular}
[c]{c}%
$2\times10^{-1}<C_{\phi}<1$\\
$\ \ \ \ \ \ \ \ \ \ \ \ 10^{-6}<C_{\phi}<10^{-2}$\\
\ \ $\ \ \ \ 10^{-11}<C_{\phi}<1$%
\end{tabular}
\\\hline\hline
$\Gamma>3H$ & Logamediate $a(t)=e^{A(\ln t)^{\lambda}}$ &
\begin{tabular}
[c]{c}%
$\ m=1$\\
$m=0$\\
$\ m=-1$%
\end{tabular}
&
\begin{tabular}
[c]{c}%
$\ \ \ \ \ \ 7\times10^{-2}<C_{\phi}<10^{-1}$\\
$\ \ \ \ \ \ \ $The model does not work\\
\ \ $\ \ \ \ \ $The model does not work
\end{tabular}
\\\hline\hline
\end{tabular}
\caption{Results for the constraints on the parameter $C_\phi$ in
the strong regime. }
\end{table}

Table II  indicates the constraints on the parameter $C_\phi$ in
the strong regime  and different choices of the parameter $m$, for
a general form for the dissipative coefficient
$\Gamma(T,\phi)=C_\phi\,T^{m}/\phi^{m-1}$, in the context of warm
intermediate and logamediate inflationary universe models.

\section{Interpolation between the weak and strong decays  \label{pp}}

Given that the ratio $R=\Gamma/3H$ will also evolve during
inflation, we may also hace models where we start  with the weak
regime ($R < 1$) but end in the strong regime ($R
> 1$).

In the following, we will analyze the intermediate and logamediate
models in the context of the interpolation between the weak and
strong decays only for the value $m=1$. For the values $m=0$ and
$m=-1$, we cannot find analytical solutions for the dissipation
coefficient given by  Eq.(\ref{G1}). For this reason, we will restrict
ourselves to the case $m=1$.

\begin{figure}[th]

\includegraphics[width=5.0 in,angle=0,clip=true]{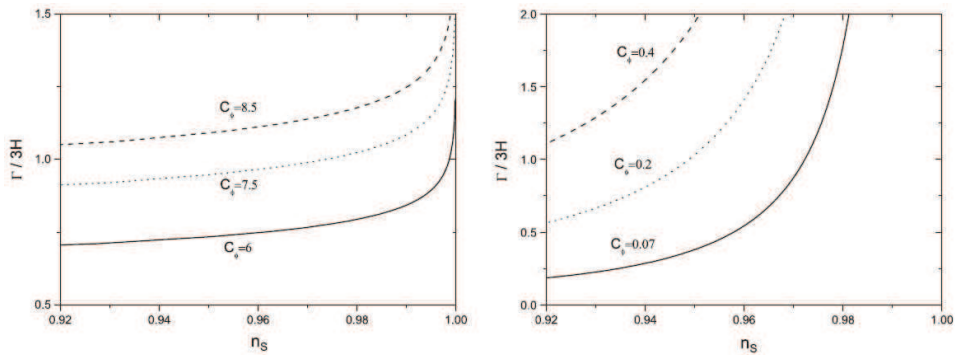}
{\hspace{0.3cm}}
\caption{The panel on the left shows the evolution of the ratio
$R=\Gamma/3H$ versus the primordial tilt $n_s$ in the
interpolation between the weak and strong decays for the
warm intermediate model, and the right panel shows the evolution
of $R=\Gamma/3H$ versus $ n_s$ for the warm logamediate model. In
both panels we use three different values of the parameter
$C_\phi$, and $m=1$, $\kappa=1$ and
$C_\gamma=70$. Also, in the left panel we use $A=0.4$ and
$f=0.9$, and in the right panel we use $A=0.002$ and
$\lambda=4$.\label{fig8}}
\end{figure}

From Eq.(\ref{G1})  and considering the case $m=1$ together with
the intermediate and logamediate models, we obtain a real
solution for the dissipation coefficient $\Gamma$  as a function
of  time, i.e., $\Gamma=\Gamma(t)$, for each of the models. In
Fig.\ref{fig8}, we show the dependence of the ratio
$R=\Gamma/3H$ and the tensor-to-scalar ratio $r$ on the
primordial tilt $n_s$ for the special case in which we fix
$m=1$ in the warm weak strong dissipative regime for the
intermediate and logamediate models. In both panels we use
three different values of the parameter $C_\phi$. The upper panel
shows the evolution of the rate $R=\Gamma/3H$ during the warm
intermediate model and we verify the evolution of the rate $R$ from
the weak and strong decays. Here we observe that the value
$C_\phi<10$ is well supported from the interpolation $R<1$ and
$R>1$. The lower panel shows the evolution of the rate
$R=\Gamma/3H$ during the warm logamediate scenario, and as before
we verify  the evolution of the rate $R$. Also, we note that the
value $C_\phi<1$ is well supported from the interpolation $R<1$
and $R>1$. For values of $C_\phi<10^{-2}$ we find that the model
evolves according to the the weak dissipative regime, i.e., $R<1$.

In the following, we will describe the scalar and tensor
perturbations  for our warm model during the interpolation between
the weak and strong decays. For a standard scalar field the
density perturbation between the weak and strong regime can be
written as \cite{warm}

\begin{equation}
{\cal{P}_{\cal{R}}}=\frac{\sqrt{3\,\pi}}{2}\,\frac{H^3\,T}{\dot{\phi}^2}\,(1+R)^{1/2}.
\end{equation}
By using Eqs.(\ref{inf3}) and (\ref{rh-1}) the density
perturbation can be written as
\begin{equation}
{\cal{P}_{\cal{R}}}=
\sqrt{\pi}\,H^3\,\left[\frac{3^3\,\kappa^3}{2^9}\,\frac{R\,(1+R)^{5}}{C_\gamma\,(-\dot{H})^3}\right]^{1/4},
\end{equation}
and the tensor-to-scalar ratio becomes
\begin{equation}
r\simeq\frac{1}{H}\,\left[\frac{2^{13}}{3^3\,\pi^{10}\,\kappa}\,\frac{C_\gamma\,(-\dot{H})^3}{R\,(1+R)^{5}}\right]^{1/4}.
\end{equation}

\begin{figure}[th]
\includegraphics[width=2.5 in,angle=0,clip=true]{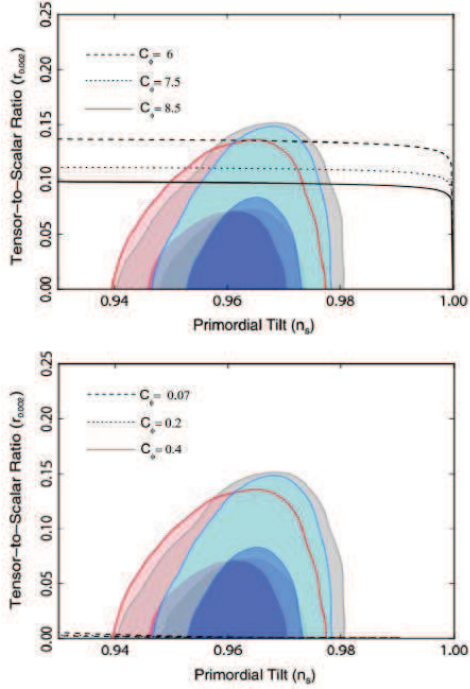}
{\hspace{0.3cm}}
\caption{The upper and lower panels show the evolution of the of
the tensor-to-scalar ratio r versus $ n_s$ in the interpolation
between the weak and strong decays for the
warm intermediate and warm logamediate models, respectively. As before, in both
we use three different values of the parameter
$C_\phi$, and $m=1$, $\kappa=1$, and
$C_\gamma=70$. In both panels  we show the two-dimensional
marginalized constraints (68$\%$ and 95$\%$ C.L.) on the
inflationary parameters $r$ and $n_s$ derived from Planck
\cite{Planck}. As before, in the upper panel we use $A=0.4$
and $f=0.9$, and in the lower panel we use $A=0.002$ and
$\lambda=4$.\label{fig9}}
\end{figure}

In Fig.\ref{fig9} we show the dependence of the tensor-to-scalar
ratio $r$ on the spectral index $n_s$ during the
interpolation between the weak and strong dissipative regimes for
the case $m=1$. In both  panels we show the two-dimensional
marginalized constraints (68$\%$ and 95$\%$ C.L.) on the inflationary
parameters $r$ and $n_s$ from Planck data\cite{Planck} for the
warm intermediate  and  warm logamediate models.  As before, in
both panels we use three different values of the parameter
$C_\phi$.

From the upper panel, in the  which the scale factor grows with the
intermediate expansion, we note that for the values of $C_\phi>1$
the model is well supported by the Planck data. Here we observe
that the curves $r=r(n_S)$ for Planck data enter the 95$\%$
confidence region only. From the lower panel, in the  which the
scale factor growths with the logamediate expansion, we note that for
the values of $C_\phi>10^{-2}$ the model is well
supported by the Planck data. Also, we note that in this model
the curves $r=r(n_s)\sim 0$. In this form, for the value $m=1$, we
get $1<C_\phi<10$ for the warm intermediate model and
$10^{-2}<C_\phi<1$ for the warm logamediate model.

\section{Conclusions \label{conclu}}

In this paper we have investigated the intermediate and
logamediate inflationary model in the context of warm inflation.
In the slow-roll approximation we have found solutions of the
Friedmann equations for a flat universe containing  a standard
scalar field in the weak and strong regime for a general form of
the dissipative coefficient
$\Gamma(T,\phi)=C_\phi\,T^{m}/\phi^{m-1}$. In particular, we
studied the values $m=1$, $m=0$, and $m=-1$. From the
warm intermediate and logamediate inflationary models, we have
obtained explicit expressions for the corresponding scalar
potential, power spectrum of the curvature perturbations, tensor-to-scalar ratio and scalar spectrum index.

For both regimes, we have considered the constraints on the
parameters of the models from the WMAP nine-year data and Planck
data. Here we have taken the constraint $r$-$n_s$ plane at lowest
order in the slow-roll approximation. Also, we found a
constraint for the value of $C_\phi$ from the weak (strong) regime
$R=\Gamma/3H< 1$ ( $R=\Gamma/3H> 1$ ). It is interesting to note
that in general we have obtained an upper bound  for the
parameter $C_\phi$ from this condition. Also,  we noted that when
we decrease the value of the parameter $m$ the value of the
parameter $C_\phi$ also decreases. Our results are summarized in
Tables I and Table II, respectively.

During the interpolation between weak and strong decays, we
analyzed the constraints on the parameters from Planck data only
for the case $m=1$ in the warm intermediate and the
warm logamediate models.

\begin{acknowledgments}
R.H. was supported by COMISION NACIONAL DE CIENCIAS Y TECNOLOGIA
through FONDECYT grant N$^{0}$ 1130628 and by DI-PUCV   N$^{0}$
123724. M.O. was supported by PUCV. N.V. was supported by Proyecto
Beca-Doctoral CONICYT N$^0$ 21100261.
\end{acknowledgments}

%\\\\\\\\\\\\\\\\\\\\\\\\\\\\\\\\\\\\\\\\\\\\\\\\\\\\\\\\\\\\\\\\\\\\\\\

\end{document}